\documentclass[a4paper,singlecoulomb,10pt]{article}
\usepackage{epsfig,amsmath,amssymb}
\usepackage[colorlinks=true,linktocpage=true,linkcolor=blue,citecolor=blue]{hyperref}
\usepackage{cite}
\usepackage{appendix}
\usepackage{graphicx}
\usepackage{dcolumn}
\usepackage{bm}
\usepackage{subfig}
\usepackage{float}
\usepackage{tabularx}
\usepackage{hhline}
\usepackage{color}
\usepackage{latexsym}
\usepackage{caption}

\begin{document}
\large
\title{\bf  Magnetic field-induced anisotropic interaction in heavy quark bound 
states}
\author{Salman Ahamad Khan\footnote{skhan@ph.iitr.ac.in}, Binoy Krishna
Patra\footnote{binoy@ph.iitr.ac.in} and Mujeeb Hasan\footnote{
mhasan@ph.iitr.ac.in}\vspace{0.3in} \\
Department of Physics, \\
Indian Institute of Technology Roorkee, Roorkee 247667, India}
\date{}
\maketitle
\vskip 0.01in

\begin{abstract}
In this article we have investigated how a strong magnetic field (${\bf B}$)
could decipher the anisotropic interaction in heavy quark ($Q$) and antiquark
($\bar Q$) bound states through the perturbative thermal QCD 
in real-time formalism. So we thermalize the Schwinger propagator for 
quarks in the lowest Landau level and the Feynman propagator
for gluons to calculate the gluon self-energy up to one loop
for massless flavours. For the quark-loop contribution to
the self-energy, the medium does not have any temperature correction and the
vacuum term gives rise an anisotropic term whereas the gluon-loop
yields the temperature correction. This finding in quark-loop
contribution corroborates the equivalence of a massless QED in (1+1)-dimension 
with the massless thermal QCD in strong magnetic field, which 
(quark sector) is reduced to (1+1)-dimension (longitudinal).  
This anisotropy in the self-energy is then 
being translated into the permittivity of the medium, which now
behaves like as a tensor. Thus the permittivity 
of the medium makes the $Q \bar Q$ potential anisotropic in the presence of 
strong magnetic field in the coordinate space, which 
resembles with a contemporary results found in lattice studies
\cite{Bonati:PRD89'2014,Bonati:PRD94'2016}. As a matter of fact, the 
potential for $Q \bar Q$-pairs 
aligned transverse to ${\bf B}$ is more attractive than the parallel 
alignment. However, the potential is always more attractive compared
to the absence of ${\bf B}$ due to the softening of the electric screening 
mass. However, the (magnitude) imaginary-part of the potential
becomes smaller compared to $B=0$. We have next
investigated the effects of strong ${\bf B}$ on the binding
energies (B.E.) and thermal widths ($\Gamma$) of the ground states of 
$c \bar c$ and
$b \bar b$ in a first-order time-independent perturbation theory, 
where the binding energies gets increased and the widths 
gets decreased compared to $B =0$. The above 
medium modifications to the properties of $ Q \bar Q$ bound 
states facilitated to study their quasi-free
dissociation in the medium in a strong magnetic
field. The dissociation temperatures estimated for $J/\psi$ and $\Upsilon$ 
states quantitatively from an optimized criterion - B.E.=$\Gamma$/2 are
obtained as $1.59 \rm{T_c} $ and $2.22 \rm{T_c}$, respectively, which
are higher than the estimate in the absence of strong magnetic
field. Thus strong $B$ impedes the early dissolution of $Q \bar Q$ bound 
states.
\end{abstract}

\section{Introduction}
Lattice Quantum Chromodynamics predicts
that at extreme conditions of high temperatures and/or high
densities, quarks confined inside the hadrons get deconfined and roam
in an extended region of space (much bigger than the size of a hadron),
known as quark-gluon plasma (QGP). This novel phenomenon can be 
seen as a generic property of nonabelian gauge 
theories at high energies, celebrated as Asymptotic Freedom. It is
believed that such state of matter also existed in our present universe 
around one microsecond after the big bang, in the core of the dense stars,
in the terrestrial laboratory of ultra-relativistic heavy ion collision 
(URHIC) experiments etc. As we know from the ongoing URHIC experiments 
at Relativistic Heavy Ion Collider (RHIC) and Large
 Hadron Collider (LHC), a very strong magnetic field, 
perpendicular to the reaction plane, is produced in the 
very early stages of the collisions due to the very large relative velocity
of spectator quarks in non-central events\cite{Shovkovy,Muller:PRD89'2013}, ranging from $m_\pi^2$ ($10^{18}$ Gauss) at RHIC
\cite{Kharzeev:NPA803'2008} to $15m_\pi^2$ at LHC \cite{Skokov:IJMPA24'2009}.
Initially, it was believed  that the magnetic field decays very 
fast just after it is produced, but the life time of the 
magnetic field is elongated if the medium have finite 
value of the electrical conductivity. 
So physicists have realized that in the presence of the 
background magnetic field various physical quantities
associated with QGP will also get modified. In the recent
years various research activities are going on in which
the effects of the background magnetic field on the 
various properties of QGP have been incorporated, which in turn 
lead to many novel phenomena related to QCD, {\em viz.} 
(inverse) magnetic catalysis\cite{ Leung:PRD55'1997,
Gusynin:PRD56'1997,Haber:PRD90'2014}, 
chiral magnetic effect\cite{Fukushima:PRD78'2008,Kharzeev:NPA803'2008},
chiral vortical effect in rotating QGP \cite{Kharzeev:PRL106'2011, 
Kharzeev:PPNP88'2016}, axial magnetic effect\cite{Braguta:PRD89'2014,
Chernodub:PRB'2014}, the dilepton production rate\cite
{Tuchin:PRC88'2013,Bandyopadhyay:PRD94'2016,Sadooghi:ANP376'2017}, 
the conformal anomaly and production of soft photons
 \cite{Fayazbakhsh:PRD88'2013,Basar:PRL109'2012}, 
dispersion relation in the magnetized thermal 
QED\cite{Sadooghi:PRD92'2015}, refractive indices 
and decay constants\cite{Fayazbakhsh:PRD86'2012,
Fayazbakhsh:PRD88'2013}, thermodynamical\cite{Rath:JHEP1712'2017,Karmakar:PRD99'2019}, magnetic~\cite{Rath:EPJA55'2019} and transport properties\cite{Rath:PRD100'2019
,Kurian:PRD96'2017}. Out of many signatures of the QGP, the suppression of the 
heavy quarkonia is a very promising one. 

The heavy quark and antiquark pairs are
produced in URHICs on a very short time-scale $\sim 1/2m_Q$ ($m_Q$ is 
the mass of the heavy quark) and the pair develops into the physical 
resonances (heavy quarkonia) over a formation time and traverses the QGP
and subsequently the hot hadronic matter before leaving the interacting 
system to decay into a dilepton to be detected. So by studying the properties 
of the heavy quarkonia, we can get some understanding about the medium and
{\em vice versa}. Since the masses of the heavy quarks $m_Q$ 
are much larger than the intrinsic QCD scale
 ($\Lambda_{QCD}$), the velocity of 
the heavy quarks ($v$) is very small in the bound 
states. The $Q\bar{Q}$ pair could then be treated like a 
nonrelativistic system, possessing a hierarchy of energy scales:
$m_Q> m_{{}_Q} > m_{{}_Q} v^2$ and integrating out the successive scales 
lead to a sequence of low-energy effective field theories (EFTs), {\em viz.}
nonrelativistic QCD (NRQCD), potential nonrelativistic QCD(pNRQCD) etc. 
For example, pNRQCD (by integrating out the scale $m_{{}_Q}v$) 
describes the $Q \bar Q$ bound state by a two-point function satisfying 
the Schrodinger equation through the usual Cornell potential as the 
matching coefficients in the effective Lagrangian. For the quarkonia in a 
thermal medium, pNRQCD may be generalized to finite
temperature~\cite{Brambilla:PRD78'2008}, but the hierarchy in 
thermal scales ($T>gT>g^2T$) make
the analysis complicated. For example, if the binding
energies are larger than the temperature, although the $Q \bar Q$ potential
does not get modify but quarkonia develop a finite thermal width 
due to the medium induced singlet-octet transitions~\cite{Brambilla:PRD78'2008}.
In the opposite limit (B.E$< T < gT$), the potential
acquires an imaginary component~\cite{Brambilla:PRD78'2008}. 
However, the hierarchy in scales in EFT is not always evident 
and one needs lattice techniques to test the approach, where the modification
of the quarkonium states can be studied in its spectral 
function in terms of the Euclidean meson correlation functions~
\cite{Alberico:PRD77'2008}.
At finite temperature the construction becomes worst because the 
temporal-extent is decreasing, thus inadvertently 
supports the use of potential models at finite temperature to complement the 
lattice studies.

Thus the similarity in the time scales for the production of strong magnetic 
field and the formation of heavy quarkonia motivates us
%$t_{\rm{form}}=1/E_{\rm{bin}}$, e.g. the $c\bar{c}$ form resonances at 
%$t_{c\bar{c}}\approx 0.3$ fm therefore it seems plausible 
to study the effect of the strong magnetic field on $Q\bar{Q}$ interaction.
%In the presence of the strong magnetic field limit 
%($q_f B \gg T^2 \gg m_f^2$;
%$q_f$ and $m_f$ are the electric charge and mass of flavour, respectively)
%the mass of of heavy quark is still larger than the scale related to 
%the strong magnetic field ($m_Q>>\sqrt{eB}$),
%so we can use the potential approach even in the presence of the strong 
%magnetic field. 
In the recent years the effect of the magnetic field
have been studied on the production of the heavy quarkonia 
in\cite{Guo:PLB751'2015,Machado:PRD88'2013} and 
the evolution of $J/\psi$ and the magnetic 
conversion of $\eta_c$ to the $J/\psi$ in the 
presence of the strong magnetic field 
in\cite{Marasinghe:PRC84'2011,Yang:JPG39'2012}.
Moreover, the static properties of quarkonia~\cite{cho:PRL113'2014,
Rougemont:PRD91'2015,Dudal:PRD91'2015,Sadofyev:JHEP01'2016,
Alford:PRD88'2013} as well as open heavy flavours~\cite{Machado:PRD89'2014,
Gubler:PRD93'2016,Fukushima:PRD93'2016,Das:PLB768'2017} were studied in
the presence of magnetic field. In the recent years, the properties of 
$Q\bar{Q}$ bound states in a thermal QCD medium have 
been investigated by correcting both the perturbative and nonperturbative 
part of the $Q\bar{Q}$ potential through the dielectric function in 
the real-time formalism\cite{Thakur:PRD89'2014} and later
extended to the moving medium \cite{Thakur:PRD95'2017}. 
Two of us have also recently studied the effect of the strong 
magnetic field on the static properties of $Q \bar Q$ bound states 
as well as their dissociation in a thermal QCD medium.
%The $Q\bar{Q}$ complex potential in the magnetized QCD medium has been 
%studied in\cite{Singh:PRD97'2018} also. 

There have been lattice results on the heavy-quark potential and
screening masses, both of which show novel anisotropic behaviors
between transverse and longitudinal directions with respect to the
magnetic field direction~\cite{Bonati:PRD89'2014,Bonati:PRD94'2016,Bonati:PRD95'2017}.
The anisotropic behaviours in the heavy quark potential can be viewed
in general as a manifestation of the breaking of rotational invariance
in the presence of magnetic field. In nonrelativistic quantum mechanics,
assuming the electron possessing the spin, the orientational term in the 
potential energy arises due to the 
interaction of spin magnetic moment with the external magnetic field. In
relativistic quantum mechanics, the Dirac equation in the nonrelativistic
limit manifests the aforesaid orientational term in Pauli-Schrodinger
equation. However, in abovementioned potential studies at finite
temperature~\cite{Hasan:EPJC77'2017,
hasan:NPA995'2020,Singh:PRD97'2018} based on the pertubative thermal QCD, the
magnetic field did not reveal any anisotropic nature in $Q \bar Q$
interaction, {\em like} in aforesaid lattice studies. Therefore,  our
aim is to uncover the tensorial 
(anisotropic) part in $Q\bar{Q}$ interaction by an 
an external magnetic potential. In fact, we have found and the potential 
becomes anisotropic and depends on the relative orientation of the quark 
pairs with respect to the direction of the magnetic field.  

Initially the dissociation process of the heavy quarkonia was understood in 
terms of color screening. However, the broadening of the widths of the 
resonances is nowadays considered as the main reason behind the dissociation
and arises either due to the inelastic parton scattering process mediated by
the spacelike gluons known as Landau damping \cite{Laine:JHEP03'2007} or due to 
the gluo-dissociation process in which the color singlet state 
undergoes into a color octet state by a hard thermal gluon 
\cite{Brambilla:JHEP1305'2013}, photo-gluon dissociation. 
However, when the temperature of the medium is smaller than the 
binding energy of the particular resonance the later 
process become dominant. Thus, due to the broadening in the medium,
the quarkonium resonances is dissociated at smaller temperatures
with respect to the dissociation due to the color screening alone. 
We therefore first want to see the effects of strong
magnetic field on the screening and Landau damping, which in turn
gives the modified binding energies and widths of the resonance states, 
respectively.  In the framework of potential model studies, the aforesaid
studies are made possible 
by deriving the real- and imaginary-parts of potential in one-loop thermal 
QCD in a strong magnetic field. 

This paper is organized as follows: In Section 2, we have revisited the 
heavy quarkonia in isotropic thermal QCD to make
a baseline for our work in magnetic field in Section 3. As we know that the
strong magnetic field generically causes a momentum anisotropy, so we
have calculated the quark contribution to the gluon self-energy 
in strong $B$ at subsection 3.1 through a novel 
diagrammatic approach, by using Schwinger
 propagator. Thus results of subsection 3.1 facilitates to 
calculate the real-part of the complex permittivity 
from the static limit of resummed  propagator, thus the real-part of 
an anisotropic heavy quark potential is obtained in strong magnetic field.
Similarly we obtain the imaginary-part of the medium modified potential.
In Section 4, we have studied how the properties of
the charmonium and bottomonium ground states get affected by the
strong magnetic field, which in turn explore the dissociation of
the aforesaid states. We conclude in Section 5.

\section{Heavy Quarkonia in the absence of magnetic field}
%expansion-induced anisotropic hot QCD medium} 
The inter-quark Cornell potential between $Q$ and $\bar{Q}$ in vacuum ($T=0$),
is
\begin{eqnarray}
V(r;T=0)=-\frac{\alpha}{r}+\sigma r,
\end{eqnarray}
where $\alpha$ and $\sigma$ are the phenomenological
constants, to be fitted to reproduce the ground state
spectroscopy of heavy quarks bound states after including the
spin-dependent term in the potential.
The medium modification to the potential in the momentum space is obtained 
by the dielectric permittivity, $\epsilon$ {(\bf k)} of the medium as
\begin{eqnarray}
\tilde{V}(\mathbf k)=\frac{V(\mathbf k)}{\epsilon(\mathbf k)},
\label{mediumv}
\end{eqnarray} 
where $\epsilon(\mathbf k)$ encodes the properties of the deconfined
medium. $V (\mathbf k)$ is the Fourier transform (FT) of the vacuum 
potential, where the FT of the linear term needs to regularize 
properly. Both terms are regulated by multiplying first by a exponentially 
damping factor and then switching off after the FT is 
evaluated. 
%We assume r as distribution ($r \rightarrow 
%r\exp(-\gamma r)$). 
Thus the FT of $V(r;T=0)$ becomes 
\begin{eqnarray}
V(\mathbf{k})=-\sqrt{\frac{2}{\pi}}\frac{\alpha}{\mathbf{k}^2}-
\frac{4\sigma}{\sqrt{2\pi}\mathbf{k}^4}.
\end{eqnarray}
Finally, the medium modification to the potential in the co-ordinate space 
yields after taking the inverse FT
\begin{eqnarray}\label{inverse}
V (r;T)=\frac{1}{(2\pi)^{3/2}}\int {d^3\textbf{k}}~\frac{V(\mathbf{k})}
{\epsilon(\mathbf{k};T)}(e^{i\bf{k.r}}-1),
\end{eqnarray}
Let us first revisit the dielectric permittivity, $\epsilon (\mathbf{k})$ in an 
isotropic hot QCD medium through the
diagrammatic approach in real-time formalism, to make a baseline to
compare our work in strong magnetic field.

\subsection{Dielectric Permittivity in isotropic thermal QCD}
In the real-time formalism of thermal field theory, the form of propagator 
generically develops a matrix structure
\begin{eqnarray}
 \label{2a4}
  D^0  =  \left (\begin{array}{cc} D_{11}^0 & D_{12}^0\\
                             D_{21}^0 & D_{22}^0\\
            \end{array} \right )~,
\end{eqnarray}
whose $ij$-elements can be equivalently written in terms of 
retarded (R), advanced (A) and symmetric (S) propagators in Keldysh 
representation, 
\begin{eqnarray}
\label{2a7}
   \Delta_R^0 = D_{11}^0 - D_{12}^0 ~,~\Delta_A^0 = D_{11}^0 - 
   D_{21}^0 ~,~
   \Delta_S^0 = D_{11}^0 + D_{22}^0~.
\end{eqnarray}
Similar representation holds good for the self-energy matrix, so the 
retarded, advanced and symmetric self-energy can be written from the 
components of the self-energy matrix [$\Pi_{ij}$] can be written as
\begin{eqnarray}
\label{2a6}
   \Pi_R = \Pi_{11} + \Pi_{12}~,~ \Pi_A = \Pi_{11} + \Pi_{21} ~,~
   \Pi_S = \Pi_{11} + \Pi_{22}~.
\end{eqnarray}

Then the full or resummed retarded (advanced) and symmetric propagators 
can be obtained by resumming the above respective propagators through 
the Dyson-Schwinger equation,
\begin{eqnarray}
 \Delta_{R,A}&=&\Delta_{R,A}^0+\Delta_{R,A}^0\Pi_{R,A}{\Delta}_{R,A}~, 
\label{retarded11}\\
 {\Delta}_{S}&=&\Delta_{S}^0+\Delta_{R}^0\Pi _R{\Delta}_{S}+\Delta_S^0\Pi_{A} 
{\Delta}_{A}+\Delta_{R}^0\Pi _{S}{\Delta}_{A}~, \label{symmetric11}
\end{eqnarray}
where the symmetric one \eqref{symmetric11} can be further expressed in 
terms of retarded and advanced ones 
\begin{eqnarray}
\Delta_{S}(K)&=& (1+2f_B)\, \mbox{sgn}(k_0)\,
\left[\Delta_{R}(K)-{\Delta}_{A}(K)\right]
\nonumber \\
&&+\Delta_{R}(K)\,\left[\Pi_S (K)-(1+2f_B)\, \mbox{sgn}(k_0)\, [\Pi_R(K)-
\Pi _A(K)]\right] \,  \Delta_A(K)~. \label{2b8}
\end{eqnarray}

However, for our problem on the static potential, only the longitudinal 
component of the resummed propagators will suffice our purpose, so
the above resummed propagators \eqref{retarded11} and \eqref{symmetric11}
 will be specifically
\begin{eqnarray}
&& \Delta^{L}_{R,A}=\Delta^{L (0)}_{R,A}+\Delta^{L (0)}_{R,A}
\Pi^{L}_{R,A} \Delta^{L}_{R,A}~,\\ \label{3b2}
&&\Delta^{L}_{S}=\Delta^{L(0)}_{S}+\Delta^{L(0)}_{R} 
\Pi^{L}_R \Delta^{L}_{S(0)}
+\Delta^{L(0)}_S\Pi^{L}_A {\Delta}^{L(0)}_{A}+
\Delta^{L(0)}_{R}\Pi^{L}_S \Delta^{L(0)}_{A}~,
\label{symmetric}
\end{eqnarray}
which can be written for gluons ($\Delta=D^{\mu \nu}$, say) in Breit-Wigner
form 
\begin{eqnarray}
D^{L}_{R,A}(K)&=&\frac{1}{\mathbf{k}^2-\text{Re} \Pi^{L}_{R}(K) 
\mp i \text{Im} \Pi^{L}_{R}(K)},\label{lon_ret} \\
D^{L}_{S}(K)&=&\frac{2i~\text{Im} \Pi^{L}_{R}(K)
(1+2n_{B}(k_0)){\rm sgn} (k_0)}
{\big[\mathbf{k}^2-\text{Re} \Pi^{L}_{R}(K)\big]^2+\big[\text{Im} \Pi^{L}_{R}
(K)\big]^2},
\label{lon_sym}
\end{eqnarray}
wherein the relations between retarded and advanced self-energies for 
both real and imaginary parts have been used
\begin{eqnarray*}
\text{Re}\Pi^{L}_{R}(K) &=&\text{Re}\Pi^{L}_{A}
(K),\nonumber\\
\text{Im} \Pi^{L}_{R}(K)&=&-\text{Im} \Pi^{L}_{A}(K).
\end{eqnarray*}

The resummed retarded (or advanced) and symmetric propagators
can be inverted to obtain the (real and imaginary parts) elements of 
full propagator matrix, {\em namely} the $11$-element
\begin{eqnarray}
\text{Re}~D^{L}_{11}(K)&=&\text{Re}~D^{L}_{R} (K), 
\label{real_propagator}\\
\text{Im}~D^{L}_{11}(K)&=&\text{Im}~\frac{D^{L}_{S} (K)}{2}.
\label{imaginary_propagator}
\end{eqnarray}
The linear response theory gives the connection between the dielectric 
permittivity and the static limit of the $11$-component of resummed gluon 
propagator by 
\begin{eqnarray}
\frac{1}{\epsilon ({\mathbf k})}=\displaystyle
{\lim_{k_0 \rightarrow 0}}{\mathbf k}^{2}D_{11}^{L}(\rm k_{0}, 
\textbf{k}).
\label{dielectric}
\end{eqnarray}
Thus the real and imaginary parts of 11-component give
the respective components of the permittivity
\begin{eqnarray}
\frac{1}{{\rm Re}~\epsilon ({\bf k})}&=&\displaystyle
{\lim_{k_0 \rightarrow 0}}{\bf k}^{2}~{\rm Re}~D_R^{L}(\rm k_{0}, 
\textbf{k}),\\
\frac{1}{{\rm Im}~\epsilon ({\bf k})}&=&\displaystyle
{\lim_{k_0 \rightarrow 0}}{\bf k}^{2}~\frac{{\rm Im}~D_S^{L}(\rm k_{0}, 
\textbf{k})}{2}.
\label{dielectric}
\end{eqnarray}

We will now calculate the gluon self-energy to resum the propagators.
Let us first begin with the form of gluon self-energy tensor ($\Pi^{\mu \nu}$)
in vacuum, which could be written as 
a linear combination of the metric tensor, $g^{\mu \nu}$ and 
$K^\mu K^\nu$ (with the only four-vector available), 
\begin{eqnarray}
\Pi^{\mu\nu}(K)= \Big(g^{\mu\nu}-\frac{K^\mu K^\nu}{K^2}\Big)
\Pi (K^2) \equiv P^{\mu\nu} \Pi (K^2)
~,\end{eqnarray}
where the only (projection) tensorial basis, $P^{\mu \nu}$ satisfies the 
four-dimensional transversality condition 
\begin{eqnarray}
K_{\mu}P^{\mu\nu}=0,
\end{eqnarray}
with the additional relation
\begin{eqnarray*}
P^{\mu\rho}P_{\rho\nu}=P^{\mu}_{~\nu}. 
\end{eqnarray*}
The above scalar function, $\Pi (K^2)$ is known 
as the structure factor (self-energy), which depends on the Lorentz invariant 
quantity $K^2$.

Now bring the vacuum with the contact of a heat reservoir, which
defines a local rest frame with four velocity, 
$u^\mu = (1,0,0,0)$ and hence breaks the Lorentz symmetry
O$(1,3)$ of the vacuum into an O$(3)$ rotational symmetry. Hence
a larger basis is necessary 
%by including tensor structures 
%such as $u^\mu u^\nu$ and $k^\mu u^\nu+k^\nu u^\mu$. 
and conveniently two orthogonal tensorial basis, $P^{\mu \nu}_T$ and 
$P^{\mu \nu}_L$ have been adopted compatible for the physical degrees of 
freedom to express the tensor~\cite{Braaten:PRL64'1990,
Kobes:PRD45'1992} at finite temperature 
\begin{eqnarray}\label{g.s.e.(T)}
\Pi^{\mu\nu}(\rm k_0,\mathbf{k})=\Pi_T (\rm k_0,\mathbf{k})P_T^{\mu\nu}
+ \Pi_L (k_0,\mathbf{k})P_L^{\mu\nu}
~.\end{eqnarray}
The forms of the basis are constructed as
\begin{eqnarray}
P_T^{\mu\nu} &=& -g^{\mu\nu}+\frac{\rm k^0}{\mathbf{k}^2}\left(K^\mu u^\nu+u^\mu K^\nu\right)
-\frac{1}{\mathbf{k}^2}\left(K^\mu K^\nu+K^2u^\mu u^\nu\right),\label{transverse p.} \\ 
P_L^{\mu\nu} &=& -\frac{\rm k^0}{\mathbf{k}^2}\left(K^\mu u^\nu+u^\mu K^\nu\right)+\frac{1}
{\mathbf{k}^2}\left(\frac{(\rm k^0)^2}{\rm k^2} K^\mu K^\nu+ K^2u^\mu u^\nu\right)
\label{longitudinal p.}
,\end{eqnarray}
to satisfy the 4-dimensional transversality condition
\begin{eqnarray*}
K_{\mu}P^{\mu\nu}_T=K_{\mu}P^{\mu\nu}_L=0~. 
\end{eqnarray*}
In addition, they satisfy the following properties
\begin{eqnarray*}
P^{\mu\rho}_T P^{T}_{\rho\nu}&=&-P^{\mu}_{T~\nu}~,\\
P^{\mu\rho}_L P^{L}_{\rho\nu}&=&-P^{\mu}_{L~\nu}~,\\
P^{\mu\rho}_T P^{L}_{\rho\nu}&=&0~,
\end{eqnarray*}
where the subscripts $T$ and $L$ label the transverse and longitudinal
modes, respectively with respect to the three-momentum ($\mathbf{k}$)
and is justified by the dot products
\begin{eqnarray}
\rm k_iP^{ij}_T&=&0~,\label{trasverse_condition}\\
\rm k_iP^{ij}_L&=&-\frac{\rm (k^0)^2k^j}{K^2}\label{longitudinal_condition}.
\end{eqnarray}
The structure factors, $\Pi_T$ and $\Pi_L$ are then called transverse
and longitudinal components of self-energy tensor, respectively, which depend 
in the rest frame of the medium on both energy, $\rm k^0$ (=$K.u$)
and $|\bf k|$ (=${\rm k}= {({(K.u)}^2 - K^2)}^{1/2}$ separately due to the lack
of Lorentz invariance at finite temperature. They are calculated in 
Hard Thermal Loop (HTL) approximation with the temperature as the hard scale 
for loop momentum, however, $\Pi_T$ vanishes in the static limit.

In real-time formalism, the longitudinal component of retarded/advanced 
gluon self-energy tensor had been calculated~\cite{Weldon:PRD26'1982}
in HTL perturbation theory\footnote{$+i\epsilon $ ($ -i\epsilon $) 
prescription is for the retarded (advanced) self-energy} 
\begin{eqnarray}
\Pi^{\rm L}_{\rm R,A} ( K)=m_D^2 (T) \left(\frac{\rm k_0}{2{\rm k}}\ln\frac
{ \rm{k}_0+{k} \pm i\epsilon}{\rm k_{0}-{k} \pm i\epsilon}-1
\right)~,
\label{self_energy_gluon}
\end{eqnarray}
where $m_D^2 (T)$ is the leading-order result of the screening mass 
(also known as Debye mass) for an thermal QCD medium~\cite{Shuryak:ZETF`'1978}
and is given by
\begin{equation}
m_D^2 (T) =\left(\frac{N_c}{3}+\frac{N_f}{6}\right) g^2 T^2.
\end{equation}
Here $g$ is the running strong coupling and its one-loop expression
is given by~\cite{Laine:JHEP03'2005}
\begin{eqnarray}
\alpha (T)=\frac{g^2(T)}{4\pi}=\frac{6\pi}{(33-2N_f) \ln(\frac{Q}{\Lambda_{QCD}})}~,
\end{eqnarray}
with $N_f$ is the number of flavour (we take 3 massless
flavours) and $\Lambda_{QCD}$ is scale ($\sim 0.200$ GeV) of 
QCD. The scale, $Q$ is set at $2\pi T$.

The real- and imaginary parts of retarded self energy can thus be extracted 
as
\begin{eqnarray}
{\rm{Re}}~\Pi^{\rm L}_{\rm R,A} (\rm k_0, \bf k) 
&=&m_D^2 \left(\frac{k_0}{2\rm k}\ln{|\frac{\rm k_0+k}{\rm k_0-k}|}-1\right), \\
{\rm{Im}}~\Pi^L_{\rm R,A}(\rm k_0,  \bf k)& =& 
-\pi m_D^2\frac{{\rm k_0}}{2 \rm k}, 
\end{eqnarray}
which could then help to evaluate the resummed retarded (or advanced) and 
symmetric propagator from the Briet-Wigner formulas \eqref{lon_ret} 
and \eqref{lon_sym}, 
respectively and the real and imaginary-parts of the respective 
propagators are given by 
\begin{eqnarray}\label{ret11}
{\rm{Re}}~D^{L}_{R,A } (\rm k_0,  \bf k)&=&\frac{\mathbf{k}^2-m_D^2
\left(\frac{\rm{k_0}}{2k}\ln{|\frac{\rm k_0+k}{\rm k_0-k}|}-1\right)}
{\left[\mathbf{k}^2-m_D^2\left(\frac{\rm k_0}{2\rm k}\ln{|\frac{\rm k_0+k}{\rm k_0-k}|}-1 \right)\right]^2+ \left(\frac{\pi m_D^2\rm{k}_0}{2k}\right)^2},\\
{\rm Im}~D^{L}_{\rm S} (\rm k_0, \bf k)&=&-\frac{2Tm_D^2\pi}
{\rm k\left [\left(\mathbf{k}^2-m_D^2\left(\frac{\rm k_0}{2\rm k}
\ln{|\frac{\rm k_0+k}{\rm k_0-k}|} -1 \right)\right)^2+
\left(\frac{\pi m_D^2\rm{k_0}}{2\rm k}\right)^2\right]},
\label{sym11}
\end{eqnarray}
 where {\rm k} is $|\mathbf{k}|$.

%\begin{eqnarray}
%{\rm{Re}}~D^{L}_{R,A (\rm iso)} (\rm k_0,  \bf k)& =& \frac{1}{\bf k^2+
%m_{D}^2 }\\
%{\rm Im}~D^{L}_{\rm S(iso)} (\rm k_0, \bf k) &=& -\frac{2\pi m_D^2 T}{\bf k
%%({\bf k}^2+m_D^2)^2}~.
%\label{resummed_symmetric_gluon}
%\end{eqnarray}
Using the equations \eqref{ret11} and \eqref{sym11}, the above propagators 
in the static limit give the real- and 
imaginary-part of the (complex) dielectric permittivity in a thermal QCD 
medium 
\begin{eqnarray}
\frac{1}{ {\rm Re}~\epsilon (\bf k)} &=&\frac{\mathbf k^2}
{\mathbf k^2+m_{D}^2}
\label{real_dielectric},\\
\frac{1}{{\rm Im}~\epsilon (\bf k)} &=& 
-\pi Tm_D^2\frac{\bf{k}^2}{\rm k(\mathbf{k}^2+m_D^2)^2},
\label{imaginary_dielectric}
\end{eqnarray}
respectively. 

\subsection{Medium modification to $Q$-$\bar Q$ potential in a thermal
QCD medium}
By substituting the dielectric permittivities in \eqref{mediumv}, we could 
obtain the complex inter-quark potential in the isotropic hot QCD medium 
in the coordinate space, whose real-part is ($\hat{r}=rm_{{}_D}$)
\begin{eqnarray}
\small{
{\rm Re}~V (\hat{r};T)
=\left(\frac{2\sigma}{m_D(T)}-\alpha
m_D(T)\right)\frac{e^{-\hat{r}}}{\hat{r}}-\frac{2\sigma}
{m_D(T)\hat{r}}+\frac{2\sigma}{m_D(T)}-\alpha m_D(T).
\label{mod_pot}
}
\end{eqnarray}
and the imaginary-part is 
\begin{eqnarray}
{\rm Im}~V (\hat{r},T) =-\alpha T \phi_0 (\hat{r})-\frac{2\sigma T}{m_D^2} 
\Psi_0 (\hat{r}),
\label{imaiso}
\end{eqnarray}
where the following functions are,
\begin{eqnarray*}
\phi_ 0(\hat{r})&=&\frac{-\hat{r}^2}{9}(-4+3\gamma_ E+3\log{\hat{r}}),\\
\Psi_ 0(\hat{r})&=&\frac{\hat{r}^2}{6}+\frac{(-107+60
\gamma_ E+60\log{\hat{r}})\hat{r}^4}{3600}.
\end{eqnarray*}

%%%%%%%%%%%%%%%%%%%%%%%%%%%%%%%%%%%%%%%%%%%%%%%%%%%%%%%%%%%%%%%%%%%%
\section{Heavy Quarkonia in the presence of strong magnetic field}
%%%%%%%%%%%%%%%%%%%%%%%%%%%%%%%%%%%%%%%%%%%%%%%%%%%%%%%%%%%%%%%%%%%%
In the presence of the magnetic field, only quarks, being
electrically charged particles, are classically affected  
by the Lorentz force while gluons remain unaffected. 
%This can be further 
%evidenced by their dispersion relation quantum mechanically at finite 
%temperature in the presence and absence of the strong magnetic field. 
To be precise, in strong magnetic field limit ($|q_fB|>T^2>m_f^2$), the 
dominant scale for quarks become the magnetic field whereas for gluons 
the temperature remains the dominant scale even in the presence of strong 
magnetic field. As a result, in the presence of the strong magnetic field 
quarks and gluons are treated on different footing, hence the structure 
functions get decomposed into quark ($q$) and gluon ($g$) components. More
specifically, the abovementioned structure functions, $\Pi_L$ and $\Pi_T$
in the absence of magnetic field will now be ascribed to gluons only and 
the two new structure functions, $\Pi^\parallel$ and $\Pi^\perp$ (later
in \eqref{self_B}) are to be included in the gluon self-energy tensor for
the quarks only (the notations, $\parallel$ and $\perp$ denote
the components along and transverse
to the magnetic field, respectively).
%%%%%%%%%%%%%%%%%%%%%%%%%%%%%%%%%%%%%%%%%%%%%%%%%%%%%%%%%%%%%%%%%
%%%%%%%%%%%%%%%%%%%%%%%%%%%%%%%%%%%%%%%%%%%%%%%%%%%%%%%%%%%%%%%%%
\subsection{Gluon self-energy tensor in the presence of strong
magnetic field}
%%%%%%%%%%%%%%%%%%%%%%%%%%%%%%%%%%%%%%%%%%%%%%%%%%%%%%%%%%%%%%%%%
%%%%%%%%%%%%%%%%%%%%%%%%%%%%%%%%%%%%%%%%%%%%%%%%%%%%%%%%%%%%%%%%%
In continuation with abovementioned discussion, the form of the gluon 
self-energy tensor can be  
written as~\cite{Hattori:PRD97'2018,Hattori:AP330'2013}
\begin{eqnarray}
\Pi^{\mu\nu}(K)=\Pi^ {g,T}P_T^{\mu\nu}(K)+\Pi^{g,L}P_L^{\mu\nu}(K)+
\Pi^{q,\parallel}P_{\parallel}^{\mu\nu}(K)
+\Pi^{q,\perp}P_{\perp}^{\mu\nu}(K),
\label{self_B}
\end{eqnarray}
where $\Pi^{g,L}$ and  $\Pi^{g,T}$ are the structure factors for
gluons only and the new two structure factors, $\Pi^{q,\parallel}$ 
and $\Pi^{q,\perp}$ appear for quarks only and their evaluation is
to be done from the quark-loop. 
%$P_T^{\mu\nu}$ and $P_L^{\mu\nu}$ are the projection tensors in the 
%thermal medium. 
In the presence of the strong magnetic 
field (in the direction $b^{\mu}$), the rotational invariance of 
the thermal medium is broken and a much extended
tensor basis is required and can be constructed with the help of 
vectors $K^{\mu}$, $u^{\mu}$,
$b^{\mu}$ and the tensor $g^{\mu \nu}$.
So in addition to  $P_T^{\mu\nu}$ and  $P_L^{\mu\nu}$, two more
projection tensors $P_{\parallel}^{\mu\nu}$ and $P_{\perp}^{\mu\nu}$
have been constructed  as \cite{Hattori:PRD97'2018,Hattori:AP330'2013}
\begin{eqnarray}
P_{\parallel}^{\mu\nu}&=&-\frac{\rm k^0k^z}{\rm k_{\parallel}^2}
(b^{\mu}u^{\nu}+u^{\mu}b^{\nu})+\frac{1}{\rm k_{\parallel}^2}
(({\rm k^0})^2b^{\mu}b^{\nu}+({\rm k^z})^2u^{\mu}u^{\nu}),\\
&=&-\left(g_{\parallel}^{\mu\nu}-\frac{\rm k_{\parallel}^{\mu}
k_{\parallel}^{\nu}}{\rm k_{\parallel}^2}\right),\\
P_{\perp}^{\mu\nu}&=&\frac{1}{{\rm k^2}_{\perp}}[{\rm -k^2}_{\perp}g^{\mu\nu}
+{\rm k^0}(K^{\mu}u^{\nu}+u^{\mu}K^{\nu})-{\rm k^z}(K^{\mu}b^{\nu}+b^{\mu}K^{\nu})+
{\rm k^0k^z}(b^{\mu}u^{\nu}\\
&&+u^{\mu}b^{\nu}) -K^{\mu}K^{\nu}
+({\rm k_{\perp}^2-(k^0)^2)}u^{\mu}u^{\nu}-{K^2}b^{\mu}b^{\nu}],\nonumber\\
&=&-\left(g_{\perp}^{\mu\nu}-\frac{\rm k_{\perp}^{\mu}
k_{\perp}^{\nu}}{\rm k_{\perp}^2}\right),
\end{eqnarray}
with the following notations:
\begin{eqnarray*}
u^{\mu}&=&(1,0,0,0), \quad b^{\mu}=(0,0,0,-1),\\
g_{\parallel}^{\mu\nu}&=&\text{diag}(1,0,0,-1), 
\quad g_{\perp}^{\mu\nu}=\text{diag}(0,-1,-1,0),\\
K^2&=& \rm k_{\parallel}^2- k_{\perp}^2, \quad k_{\parallel}^2=(k_0)^2-(k_z)^2,\\
 \rm k_{\perp}^2&=& \rm (k_x)^2+ (k_y)^2.
\end{eqnarray*}
The tensorial basis satisfy the following properties~
\cite{Hattori:PRD97'2018}:
\begin{eqnarray}
P_{\parallel}^{\mu \rho}P^{\parallel}_{\rho \nu}&=&-P^{\mu}_{\parallel \nu},\\
P_{\perp}^{\mu \rho}P^{\perp}_{\rho \nu}&=&-P^{\mu}_{\perp \nu},\\
P_{\parallel}^{\mu \rho}P^{\perp}_{\rho \nu}&=&
P^{\mu \rho}_{\perp}P^L_{\rho \nu}=0,\\
P_{T}^{\mu \rho}P^{\perp}_{\rho \nu}&=&P^{\mu \rho}_{\perp}P^T_{\rho \nu}=
-P^{\mu}_{\perp \nu}.
\end{eqnarray}
In the strong magnetic field, quarks are confined only in the 
lowest Landau level ($n=0$), resulting the transverse component of the 
quark momentum negligibly small ($p_{\perp}=0$). Consequently, 
$\Pi^{q,\perp}$ becomes negligible ($\Pi^{q,\perp}\approx0$)
 \cite{Hattori:PRD97'2018,Ayala:PRD98'2018}, so the longitudinal 
 component\footnote{new notation, $L^\prime$ is due to get rid of confusion
from earlier notation of longitudinal component, L in the absence of magnetic 
field} of the gluon self-energy tensor at
 finite T and strong magnetic field due to 
 quark (q) and gluon (g) loops is given by 
 \begin{eqnarray}
 \Pi^{L'}(K)=\Pi^{q,\parallel}(K)+\Pi^{g,L}(K),
\label{long_com}
\end{eqnarray}
because $P_T^{00},~P_{\perp}^{00}=0$.

First, we will calculate $\Pi^{q,\parallel}$ from the quark-loop up to 
one-loop. As we know, in strong magnetic field, only the lowest Landau level
(LLL) are populated, so the quark propagator in vacuum in the 
momentum space is restricted to the LLL~\cite{Tsai:PRD10'1974,Chyi:PRD62'2000} 
\begin{eqnarray}
iS_0(P)=\frac{(1+\gamma^ 0\gamma^ 3
\gamma^ 5)(\gamma^ 0p_0-\gamma^ 3p_z+m_f)}{p_{\parallel}^2-m_f^2+i\epsilon}
e^{-\frac{p^2_{\perp}}{|q_fB|}}.
\end{eqnarray}
Now the above vacuum quark propagator at finite temperature in real-time
formalism becomes a matrix 
\begin{equation}
S(P) =
\begin{pmatrix}
S_0(P)+n_F(p_0) (S^\ast_0 (P) -S_0(P)) & 
\sqrt{n_F(p_0) (1-n_F(p_0))} (S^\ast_0 (P) -S_0(P)) \\
-\sqrt{n_F(p_0) (1-n_F(p_0)} (S^\ast_0 (P) -S_0(P)) &
-S^\ast_0(P)+n_F(p_0) (S^\ast_0 (P) -S_0(P)) 
\end{pmatrix}~,
\label{mag_prop}
\end{equation}
whose $11$-element is 
\begin{eqnarray}
iS_{11}(P)&=&\left[\frac{1}{p_{\parallel}^2-m_f^2+i\epsilon}
+2\pi i n_F(p_0)\delta(p_{\parallel}^2-m_f^2)\right](1+\gamma^ 0\gamma^ 3
\gamma^ 5)(\gamma^ 0p_0-\gamma^ 3p_z+m_f)\nonumber\\
&&\times e^{-\frac{p^2_{\perp}}{|q_fB|}}.
\label{quark_prop}
\end{eqnarray}

Thus the 11-component of the quark-loop contribution can be written in
strong magnetic field with the above quark propagator in LLL as 
\begin{eqnarray}
\Pi_{11}^{q,\mu\nu} (K)&=&\frac{i{g^\prime}^2}{2} \sum_ f\int \frac{dp^2_{\perp}dp_{\parallel}^2}
{(2\pi)^4}{\rm{Tr}}[\gamma^ {\mu}(1+\gamma^ 0\gamma^ 3
\gamma^ 5)(\gamma^ 0p_0-\gamma^ 3p_z+m_f)\gamma^ {\nu}
(1+\gamma^ 0\gamma^ 3\gamma^ 5)\nonumber \\
&&\times (\gamma^ 0q_0-\gamma^ 3q_z+m_f)]
 \left[\frac{1}{p_{\parallel}^2-m_f^2+i\epsilon}
+2\pi i n_F(p_0)\delta(p_{\parallel}^2-m_f^2)\right]e^{-\frac
{p^2_{\perp}}{|q_fB|}}\nonumber \\
&&\times \left[\frac{1}{q_{\parallel}^2-m_f^2+i\epsilon}
+2\pi i n_F(q_0)\delta(q_{\parallel}^2-m_f^2)\right]e^{-\frac
{q^2_{\perp}}{|q_fB|}}.
\label{quark_self}
\end{eqnarray}
Here ${g^{\prime}}^2$ is the running strong coupling and 
runs with the magnetic field only because in strong magnetic limit
the magnetic field is the hard scale~\cite{Ferrer:PRD91'2015}
\begin{eqnarray}
\alpha '(eB)=\frac{g'^2(eB)}{4\pi}=\frac{1}{(\alpha^0(\mu_0))
^{-1}+\frac{11N_C}{12\pi}
\ln \left(\frac{{\rm k_z}^2+M_B^2}{\mu_0^2}\right)+\frac{1}{3\pi}
\sum_f \frac{|q_fB|}{\sigma}},
\end{eqnarray}
where  $$\alpha^0(\mu_0)=\frac{12\pi}{11N_C \ln
\left(\frac{\mu_0^2+M_B^2}{\lambda_V^2}\right)},$$
here $M_B$ is infrared mass. $\Lambda_V$ and $\mu_0$
are taken as 0.385 GeV and 1.1 GeV, respectively and ${\rm k_z}=0.1\sqrt{eB}$.

The above self-energy tensor can be further expressed in terms of trace 
tensor, $L^{\mu \nu}$
\begin{eqnarray}
\Pi_{11}^{q,\mu\nu}(K)&=&\frac{ig'^2}{2} \sum_ f\int \frac{dp^2_{\perp}dp_{\parallel}^2}
{(2\pi)^4}L^{\mu \nu}
 \left[\frac{1}{p_{\parallel}^2-m_f^2+i\epsilon}
+2\pi i n_F(p_0)\delta(p_{\parallel}^2-m_f^2)\right]e^{-\frac
{p^2_{\perp}}{|q_fB|}}\nonumber \\
&&\times \left[\frac{1}{q_{\parallel}^2-m_f^2+i\epsilon}
+2\pi i n_F(q_0)\delta(q_{\parallel}^2-m_f^2)\right]e^{-\frac
{q^2_{\perp}}{|q_fB|}},
\label{selfenergy1}
\end{eqnarray}
with
\begin{eqnarray}
L^{\mu \nu}=8~[p^{\mu}_{\parallel}q^{\nu}_{\parallel}+
p^{\nu}_{\parallel}q^{\mu}_{\parallel}-g^{\mu\nu}_
{\parallel}(p_{\parallel}^{\mu}.q_{\parallel \mu}-m_f^2)].
\end{eqnarray}
%Here $K(k_0,\textbf{k})$ is the external momentum of the gluon % and 
%$k=|\textbf{k}|$. In the strong 
The the (quark) loop momentum is factorizable into the longitudinal
and the transverse component with respect to the 
direction of the magnetic field, which is consequently translated into 
the factorization in the external momentum of self-energy tensor as 
\begin{eqnarray}
\Pi_{11}^{q,\mu\nu}(K)=\Pi_{11}^{q,\mu\nu}(\rm k_\parallel)B(\rm k_{\perp}).
\end{eqnarray}
On integrating over the transverse component of the loop momentum, we get 
\begin{eqnarray}
B({\rm k}_{\perp})=\frac{\pi |q_fB|}{2}e^{\frac{{\rm -k}_{\perp}^2}{2|q_fB|}},
\end{eqnarray}
which, in the strong magnetic field limit ($\rm k_{\perp}\approx 0$),
yields into 
 \begin{eqnarray}
 B({\rm k}_{\perp})=\frac{\pi |q_fB|}{2}.
 \label{transverse}
 \end{eqnarray}
%%%%%%%%%%%%%%%%%%%%%%%%%%%%%%%%%%%%%%%%%%%%%%%%%%%%%%%%%%%%%%%%%%%%%%%
%\subsection{Resummation of gluon propagator in strong magnetic field}
%%%%%%%%%%%%%%%%%%%%%%%%%%%%%%%%%%%%%%%%%%%%%%%%%%%%%%%%%%%%%%%%%%%%%%%
\subsubsection{Real-part of retarded self-energy}
The real-part of the retarded (or advanced) gluon self-energy can be 
obtained 
from the real-part of the 11-component of the self energy matrix as
\begin{eqnarray}
\text{Re}~\Pi_{R,A}^{\mu \nu}(K)&=&\text{Re}~\Pi_ {11}^{\mu \nu}(K),
\end{eqnarray}
which can be obtained as the sum of the quarks(q) and gluons(g) 
loop diagrams. Since the gluon-loop are directly unaffected by the
magnetic field, we are now going to calculate
the quark-loop only, which is separated into
the vacuum (\rm vac) and medium ($n$, $n^2$) contributions 
\begin{eqnarray}
\Pi_{R,A}^{q,\mu\nu}({\rm k_{\parallel}})&=&
\Pi_{R,A(\rm vac)}^{q,\mu\nu}
({\rm k_{\parallel}})+ \Pi_{R,A(n)}^{q,\mu\nu} ({\rm k_{\parallel}})+
\Pi_{R,A(n^2)}^{q,\mu\nu}(\rm k_{\parallel}),
\label{self_iso}
\end{eqnarray}
where,
\begin{eqnarray}\label{vaccume}
\Pi^{q,\mu\nu}_{R,A(\rm vac)}(\rm k_{\parallel})&=&\frac{ig'^2}{2(2\pi)^4}\int dp_0 dp_z
L^{\mu \nu}\left[\frac{1}{(p_{\parallel}^2-m_f^2+i\epsilon)
(q_{\parallel}^2-m_f^2+i\epsilon)}\right],\\
\label{medium}
\Pi ^{q,\mu\nu}_{R,A(n)}(\rm k_{\parallel})&=&-\frac{g'^2}{2(2\pi)^3}\int dp_0 dp_z
L^{\mu \nu}\left[n_F(p_0)\frac{\delta(p_{\parallel}^2-m_f^2)}
{(q_{\parallel}^2-m_f^2+i\epsilon)}+n_F(q_0)\frac{\delta(q_{\parallel}^2-m_f^2)}
{(p_{\parallel}^2-m_f^2+i\epsilon)}\right],\\
\label{medium2}
\Pi ^{q,\mu\nu}_{R,A(n^2)}(\rm k_{\parallel})&=&-\frac{ig'^2}{2(2\pi)^2}\int dp_0 dp_z
L^{\mu \nu}\left[n_F(p_0)\delta(p_{\parallel}^2-m_f^2)
n_F(q_0)\delta(q_{\parallel}^2-m_f^2)\right],
\end{eqnarray}
where the vacuum term  is calculated as 
\cite{Hasan:EPJC77'2017,Gusynin:NPB462'1996,Fukushima:PRD83'2011,Hattori:AP334'2013}
\begin{eqnarray}
{\rm Re}\Pi^{q,\mu\nu}_{R,A(\rm vac)}({\rm k_{\parallel}})=
\left(g_{\parallel}^{\mu\nu}-\frac{\rm k_{\parallel}^{\mu}
k_{\parallel}^{\nu}}{\rm k_{\parallel}^2}\right)\frac{g'^2}{2\pi ^3}
\left[\frac{2m_f^2}{\rm k_{\parallel}^2}\left(1-\frac{4m_f^2}
{\rm k_{\parallel}^2}\right)^{-1/2}\left\{\ln \frac{\left(1-\frac{4m_f^2}
{\rm k_{\parallel}^2}\right)^{1/2}-1}{\left(1-\frac{4m_f^2}
{\rm k_{\parallel}^2}\right)^{1/2}+1}\right\}+1\right].
\end{eqnarray}
After multiplying the transverse momentum dependent factor \eqref{transverse},
the real part of the longitudinal component (labeled as $\parallel$) 
of the vacuum part for the massless quarks ($m_f$=0) reduces to 
\begin{eqnarray}
{\rm Re}\Pi^{q,\parallel}_{R,A(\rm vac)}({\rm k_0,k_z})=
\frac{g'^2}{4\pi^2}\sum_ f|q_f|B\frac{\rm k_z^2}{\rm k_{\parallel}^2}.
\label{re_vac}
\end{eqnarray}

Next the real part of the longitudinal ($\parallel$) component due to the 
medium contribution having single distribution ($n$) function \eqref{medium} 
can be written as
	\begin{eqnarray}
	{\rm Re} {\Pi^{q,\parallel}_{R,A(n)} ({\rm k_\parallel)}} &=&\frac{g'^2}
	{2(2\pi)^3}\int dp_0 dp_z L^{00}\Bigg[
	n_F (p_0)\frac{\Big\{\delta(p_{0}-\omega_{p})
	+\delta(p_{0}+\omega_{p})\Big\}}{(q_{0}^2-\omega_q^2)
	(2\omega_{p})}\nonumber \\
	&+&n_F (q_0)\frac{\Big\{\delta(q_{0}-\omega_{q})
	+\delta(q_{0}+\omega_{q})\Big\}}{(p_{0}^2-\omega_p^2)
	(2\omega_{q})}\Bigg],
	\label{pn00}
	\end{eqnarray}
with the notations 
\begin{eqnarray}
L^{00}&=&8[p_{0}q_{0}+p_{z}q_{z}+m_f^{2}],\\
\omega_p&=&\sqrt{p_z^2+m_f^2},\\
\omega_q&=&\sqrt{(p_z-k_z)^2+m_f^2}.
\end{eqnarray}
In HTL approximation, the above medium contribution due to the
quark-loop for massless flavours vanishes
\begin{eqnarray}
{\rm Re}~\Pi^{q,\parallel}_{R,A(n)}({\rm k_0,k_z})=0,
\label{re_n}
\end{eqnarray}
which ought to be because in strong magnetic field limit, QCD for massless 
flavours is equivalent to two dimensional massless QED (Schwinger model),
which is not subject to any temperature/density corrections for the 
dimensional reason~\cite{Fukushima:PRD93'2016,Dolan:PRD9'1974}.

%It has been known for a long time that there is no temperature correction in 
%the massless Schwinger model. This point is shown in, e.g, [Dolan and Jackiw, 
%PhysRevD.9.3320]. An equivalent in the current case is the longitudinal 
%(1+1)-dimensional part in the lowest Landau level approximation.

Similarly the $n^2$-medium contribution \eqref{medium2} does not contribute
to the real part
\begin{eqnarray}
{\rm Re}\Pi^{q,\parallel}_{R,A(n^2)}({\rm k_0,k_z})=0,
\label{re_n2}
\end{eqnarray}
which represents the inelastic processes.

Thus the real-part of the longitudinal component of retarded
self-energy due to the quark loop simply becomes 
\begin{eqnarray}
{\rm Re}\Pi^{q,\parallel}_{R,A}({\rm k_0,k_z}) =
\frac{g'^2}{4\pi^2}\sum_ f|q_fB|\frac{\rm k_z^2}{\rm k_{\parallel}^2},
\label{iso_f}
\end{eqnarray}
which, in the static limit (${\rm k_z}={|\bf{k}|}\cos{\beta_n}$)
takes the form
\begin{eqnarray}
\text{Re}\Pi_{R,A} ^{q,\parallel}{(\rm k_0=0, \bf k\rightarrow 0)}&=&
-\frac{g'^2}{4\pi^2}\sum_ f|q_fB|\cos^ 2{\beta_ n},
\label{quark_part}
\end{eqnarray}
where $\beta_ n$ is the angle between the momentum $\bf{k}$ 
and the direction of the anisotropy $\bf{n}$ (the direction of the 
$\bf{B}$). 

Therefore, the real-part of the longitudinal component 
(denoted by $L^\prime$) of the retarded gluon self-energy 
tensor \eqref{self_B} in strong magnetic field is the sum of 
quark \eqref{quark_part} and gluon contribution 
\eqref{self_energy_gluon} in the static limit
\begin{eqnarray}
\text{Re}~\Pi_{R,A}^{L'} {(\rm k_0=0,\bf k \rightarrow 0)}&=&
\text{Re}~\Pi_{R,A}^ {g,L}({\rm k_0=0,\bf k \rightarrow 0})+
\text{Re}~\Pi_{R,A}^{q,\parallel}({\rm k_0=0,\bf k \rightarrow 0})
, \nonumber\\
&=&-N_C\frac{g^2 T^2}{3} -\frac{g'^2}{4\pi^2}\sum_ f|q_fB| \cos^2\beta_n,
\nonumber\\
&\equiv & -\left[{m_D^g}^2 (T) + {m_D^q}^2 (B) \cos^2\beta_n\right],
\label{real_self}
\end{eqnarray}
where
\begin{eqnarray}
{m_D^g}^2 (T) &=&\frac{N_C}{3} g^2T^2\\
{m_D^q}^2(B)&=&\frac{g'^2}{4\pi^2}\sum_ f|q_fB|.
\end{eqnarray}
Thus in the presence of strong magnetic field, the Debye mass 
for the massless quarks in thermal QCD acquires angular dependence.

It is worth to mention here that an angular dependence in
the Debye mass could arise from the momentum anisotropy inherited 
by a medium exhibited, where both quark- and gluon-loop contribute
secularly to the anisotropy in the self-energy~\cite{Thakur:PRD89'2014}. On
the contrary, the anisotropy manifested 
in our case could be understood physically
from the interaction of intrinsic spin (spin
magnetic moment) with the external magnetic field.
However, the magnetic field may also induce 
%a momentum anisotropy for a flavour (f) through its dispersion relation:
%\begin{eqnarray}
%\omega _{f,n}(p_L)=\sqrt{p_L^2+m_f^2+2n|q_fB|},
%\end{eqnarray}
%However, if the magnetic field is strong enough, quarks are confined in 
%the LLL ($n=0$), implying that the momentum carried by quarks becomes 
%longitudinal ({\em i.e.}  $p_T \approx 0$) and results an 
an anisotropy in the momentum distribution of quarks, $n_F (p_0)$). As a 
consequence, the quark propagator in \eqref{quark_prop} becomes anisotropic, 
which, in turn, makes the self-energy in \eqref{quark_self} anisotropic, 
which, for weak momentum anisotropy, is decomposible 
into isotropic and anisotropic components. We have found that the 
anisotropic component in the quark-loop is found to vanish for massless 
flavours (shown in Appendix \ref{anisotropy}). The vanishing
result can be understood by realizing the equivalence between the thermal
QCD in strong $B$ in the limit of massless flavours and
the massless QED in (1+1)-dimension, which does not allow to have any 
medium contribution because the momentum anisotropy discussed hereinabove 
is a medium description.
%so there will no anisotropic component of the self energy, {\em unlike} in
%expansion-driven anisotropy, where both
%isotropic and anistropic components exist in quark and gluon
%contributions to the self-energy.
%\begin{eqnarray}
%m'{_D}^2 (T,B)={m_D^g}^2(T)+{m_D^q}^2 (B) \cos^2\beta_n .
%\label{Debye_mag}
%\end{eqnarray}
%%%%%%%%%%%%%%%%%%%%%%%%%%%%%%%%%%%%%%%%%%%%%%%%%%%%%%%%%%%%%%%
\subsubsection{Imaginary-part of the retarded gluon self-energy}
The imaginary part of the retarded gluon self-energy tensor in the 
real-time formalism is given by~\cite{Kobes:NPB260'1985} 
\begin{eqnarray}
\text{Im}\Pi_ {R}^{\mu\nu}(K)=\frac{\tanh{\left(\frac{\beta {\rm k_0}}{2}\right)}}
{\varepsilon(\rm k_0)}\text{Im}\Pi_ {11}^{\mu\nu}(K),
\end{eqnarray}
which can again be obtained from the quark(q) and gluon (g) loops. 
Now we calculate the imaginary-part of the quark-loop contribution 
wherein the vacuum contribution
\eqref{vaccume} in the massless $m_f=0$ limit vanishes  
\begin{eqnarray}
\text{Im}\Pi_{R(\rm vac)}^{q,\parallel}({\rm k_0,k_z})=0,
\label{ima_vac}
\end{eqnarray}
and the medium ($n$-dependent) contribution \eqref{medium} also does not 
contribute 
\begin{eqnarray}
\text{Im}\Pi_{R(n)}^{q,\parallel}({\rm k_0,k_z})=0.
\label{ima_n}
\end{eqnarray}
The only nonvanishing contribution to the imaginary-part of the 
retarded self-energy comes out from the the $n^2$-term, which 
is calculated from the retarded current-current correlator in (1+1)-dimension
because the transverse dynamics gets decoupled from the longitudinal 
dynamics of LLL states in Ref. \cite{Fukushima:PRD93'2016} as
\begin{eqnarray}
\text{Im}\Pi_{R}^{q,\parallel}({\rm k_0,k_z})=-\frac{g'^2}{8\pi}
 \sum_ f|q_fB|~ \rm k_0\Big[\delta(k_0+k_z)+\delta(k_0-k_z)\Big],
\label{ima_self}
\end{eqnarray}
where the factor $\frac{|q_f B|}{8\pi}$ appears as the transverse 
density of states for the LLL states. 
\par
While, the imaginary part due to the gluon-loop contribution 
can be obtained from the known result~\eqref{self_energy_gluon} 
\begin{eqnarray}
\text{Im}\Pi_{R}^{g,L}({\rm k_0,\rm k})=
-\pi{m_D^g}^2 \frac{\rm k_0}{2\rm k}.
\end{eqnarray}
Therefore, the longitudinal component of the imaginary part of gluon 
self energy in the presence of strong magnetic field 
\begin{eqnarray}
\text{Im}\Pi_{R}^{L'} 
({\rm k_0, k})=-\frac{g'^2 \sum_ f|q_fB|}{8\pi}
{\rm k_0 \Big[\delta(k_0+k_z)}+\delta({\rm k_0-k_z})\Big]
-\pi{m_D^g}^2 \frac{\rm k_0}{2\rm k},
\label{ima_oo}
\end{eqnarray}
which in the static limit yields
\begin{eqnarray}
\lim_{\rm k_0 \to 0}\left[\frac{\text{Im}\Pi_{R}^{L'}(\rm k)}{\rm k_0}\right]
&=&
\lim_{\rm k_0 \to 0}\left[\frac{\text{Im}\Pi_{R}^{q,\parallel}(\rm k)}{\rm k_0}\right]+
\lim_{\rm k_0 \to 0}\left[\frac{\text{Im}\Pi_{R}^{g,L}(\rm k)}{\rm k_0}\right] \nonumber\\
&=&-\frac{g'^2}{8\pi} \sum_f|q_ fB|~
\Big[\delta({\rm k_z})+\delta({\rm -k_z})\Big] -\pi{m_D^g}^2
\frac{1}{2\rm k}.
\label{ima_totself}
\end{eqnarray}
%%%%%%%%%%%%%%%%%%%%%%%%%%%%%%%%%%%%%%%%%%%%%%%%%%%%%%%%%%%%%%%%%%%%%%%
\subsection{Resummed gluon propagator and permittivity}
%%%%%%%%%%%%%%%%%%%%%%%%%%%%%%%%%%%%%%%%%%%%%%%%%%%%%%%%%%%%%%%%%%%%%%%
Now, we are in a position to resum the retarded (or advanced) and symmetric 
propagators in a strong magnetic field. By substituting the real- and 
imaginary-part of the retarded (and advanced) self-energy from 
\eqref{real_self} and \eqref{ima_totself}, respectively, we have 
calculate the real-part of the resummed 
retarded propagator from the Briet-Wigner formula \eqref{lon_ret}
in the static limit
\begin{eqnarray}
{\rm Re}~D_{R,A}^{L'}(\rm k_0=0,\mathbf{k})&=&\frac{1}{\mathbf{k}^2+{m_D^g}^2+{m_D^q}^2
\cos^2\beta_n }.
\label{retard_prop}
\end{eqnarray} 
Similarly, the imaginary-part of the resummed symmetric 
propagator from Breit-Wigner formula \eqref{lon_sym} in the static limit
can be written as the sum of the quark- and gluon-loop contributions
\begin{eqnarray}
{\rm Im}~D_S ^{L'}(\mathbf{k})&=&{\rm Im}~D^{q,\parallel}_S 
(\mathbf{k})+ {\rm Im}~D^{g,L}_S (\mathbf{k}), \label{sym_prop}\nonumber\\
&=&-\frac{Tg'^2}{2\pi}\sum_f |q_ f B|\Big[\delta({\rm k_z)+\delta(-k_z) }\Big]
\frac{1}{({\mathbf{k}}^2+{m_D^g}^2+{m_D^q}^2\cos^2\beta_n)^2}\nonumber\\
&&-2\pi T{m^g_D}^2   \frac{1} {{\rm k}(\mathbf{k}^2+{m_D^g}^2+{m_D^q}^2\cos^2\beta_n)^2}.
\label{sym_gluon}
\end{eqnarray}

Therefore, the real and the imaginary-part of the dielectric permittivity
are obtained from the real-part of the resummed retarded and the 
imaginary-part of the symmetric 
propagators, respectively \eqref{real_propagator} and 
\eqref{imaginary_propagator}, where the real-part is 
\begin{eqnarray}
\frac{1}{{\text Re}~\epsilon(\mathbf{k};T,B)}&=&\frac{\mathbf{k}^2}
{\mathbf{k}^2+{m_D^g}^2+{m_D^q}^2\cos^2\beta_n},
\label{dielectric1}
\end{eqnarray}
and the imaginary part is written as the sum of the quark and gluon contributions 
\begin{eqnarray}
\frac{1}{\text{Im}~\epsilon(\mathbf{k};T,B)}&=&\frac{1}{\text{Im}~
\epsilon ^q(\mathbf{k};T,B)}+
\frac{1}{\text{Im}~\epsilon ^g(\mathbf{k};T,B)},
\label{ima_per}
\end{eqnarray}
with 
\begin{eqnarray}\label{quark_per}
\frac{1}{\text{Im}~\epsilon ^q(\mathbf{k};T,B)}&=&-\frac{Tg'^2}{4\pi}\sum_f |q_ f B|\Big[\delta({\rm k_z)+\delta(-k_z) }\Big]
\frac{\mathbf{k}^2}{({\mathbf{k}}^2+{m_D^g}^2+{m_D^q}^2\cos^2\beta_n)^2},\\
\frac{1}{\text{Im}~\epsilon ^g(\mathbf{k};T,B)}&=&-\pi T{m^g_D}^2 
\frac{ \mathbf{k}^2}
{{\rm k}(\mathbf{k}^2+{m_D^g}^2+{m_D^q}^2\cos^2\beta_n)^2},
\label{gluon_per}
\end{eqnarray}
%%%%%%%%%%%%%%%%%%%%%%%%%%%%%%%%%%%%%%%%%%%%%%%%%%%%%%%%%%%%%%%%%%%%%%
\subsection{Medium modification to $Q$-$\bar Q$ potential in a strong 
magnetic field}
%%%%%%%%%%%%%%%%%%%%%%%%%%%%%%%%%%%%%%%%%%%%%%%%%%%%%%%%%%%%%%%%%%%%%
We will use the real and imaginary parts of the dielectric
permittivity to find the medium modification to the real 
and imaginary part of the $Q\bar{Q}$ potential, respectively. 
%%%%%%%%%%%%%%%%%%%%%%%%%%%%%%%%%%%%%%%%%%%%%%%%%%%%%%%%%%%%%%%%%%%%%%%%%
\subsubsection{Real-part}
%%%%%%%%%%%%%%%%%%%%%%%%%%%%%%%%%%%%%%%%%%%%%%%%%%%%%%%%%%%%%%%%%%%%%%%%%
The real-part of the medium modified potential is  
\begin{eqnarray}
{\rm Re}~V(\mathbf{r};T,B)&=&\frac{1}{(2\pi)^{3/2}}\int {d^3\textbf{k}}~\frac{V(\mathbf{k})}
{{\rm Re}~\epsilon(\mathbf{k};T,B)}(e^{i\bf{k.r}}-1),\nonumber\\
&=&-\frac{\alpha}{2\pi^2}\int {d^3\textbf{k}}~\frac{1}{(\mathbf{k}^2+{m_D^g}^2+{m_D^q}^2
\cos^2\beta_n) }~
(e^{i\bf{k.r}}-1),\nonumber\\
&&-\frac{4\sigma}{(2\pi)^2}\int {d^3\textbf{k}}\frac{1}
{\mathbf{k}^2(\mathbf{k}^2+{m_D^g}^2+{m_D^q}^2
\cos^2\beta_n )}~(e^{i\bf{k.r}}-1),\nonumber \\
&\equiv&{\rm Re}~V_C(\mathbf{r};T,B)+ {\rm Re}~V_S(\mathbf{r};T,B),
\label{tot_pot}
\end{eqnarray}
where the Coulomb term is separated as 
\begin{eqnarray}
{\rm Re}~V_C(\mathbf{r};T,B)&=&-\frac{\alpha}{2\pi^2}\int\frac
{d^3\textbf{k}~(e^{i\bf{k.r}}-1)}{\mathbf{k}^2+
{\mu_{D}}^2+\frac{{m_D^q}^2}{2} \cos{2\beta_n}},\nonumber\\
&=&-\frac{\alpha}{2\pi^2}\int\frac{d^3\textbf{k}~(e^{i\bf{k.r}}-1)}{\mathbf{k}^2+
{\mu_{D}}^2}+\frac{\alpha {m_D^q}^2}{4\pi^2}\int \frac
{d^3\textbf{k}~(e^{i\bf{k.r}}-1)~\cos2\beta_n}{(\mathbf{k}^2+{\mu_{D}}^2)^2},\nonumber\\
& \equiv &{\rm Re}~V_C^{(1)}(\mathbf{r},T,B)+{\rm Re}~V_C^{(2)}(\mathbf{r},T,B),
\end{eqnarray}
with $\mu_D^2=({m_D^g}^2+\frac{{m_D^q}^2}{2})$. 

Therefore the first term in the Coulomb potential ($\hat{r}= r \mu_D$)
\begin{eqnarray}
{\rm Re}~V_C^{(1)}(r;T,B)&=&-\frac{\alpha}{2\pi^2}\int \frac
{d^3\textbf{k}}{\mathbf{k}^2+{\mu_{D}}^2}(e^{i\bf{k.r}}-1),\nonumber\\
&=&-\alpha \mu_{D} \frac{e^{-\hat{r}}}
{\hat{r}}-\alpha \mu_{D} ,
\label{alpha_iso}
\end{eqnarray}
where the nonlocal term gives the correct limit of the $V(r;T,B)$ as 
$T,B \rightarrow 0$. Such term
could arise naturally in thermal QCD from the real and 
imaginary-time correlators and from the basic computations
of the real-time static potential in thermal QCD\cite{Beraudo:NPA806'2008,Laine:JHEP03'2007}. 

For evaluating the second term, we first make the transformation
with the purpose for converting the anisotropy in the momentum space to 
the coordinate space as
\begin{eqnarray}
\cos{\beta_n}=\cos{\theta_r}\cos{\theta_{kr}}
+\sin{\theta_r}\sin{\theta_{kr}}\cos{\phi_{kr}},
\label{transfer}
\end{eqnarray}
where $\beta_n$ and $\theta_r$ are the angle between $\bf{k}$
and $\bf{n}$ (in the momentum space), $\bf{r}$ and $\bf{n}$ (in the
coordinate space), respectively. $\theta_{kr}$
and $\phi_{kr}$ are the angular variables for the vectors, ${\bf k}$ 
and ${\bf r}$, respectively, in the spherical polar coordinate system. 
Thus the second term in the Coulomb sector
\begin{eqnarray}
{\rm Re}~V_C^{(2)}(r,\theta _r;T,B)&=&
\frac{\alpha {m_D^q}^2}
{4\pi^2}\int {d^3\textbf{k}}~\frac
{(e^{i\bf{k.r}}-1)~\cos2\beta_n}{(\mathbf{k}^2+{\mu_{D}}^2)^2},
\nonumber\\
&=&-\frac{\alpha {m_D^q}^2}{\mu_{D}}
\left[\frac{e^{-\hat{r}}}{\hat{r}}\left(\frac{\hat{r}}{4}
+\frac{1}{\hat{r}}+\frac{1}{\hat{r}^2}+\frac{1}{2}\right)
-\frac{1}{\hat{r}^3}-\frac{1}{12}\right.\nonumber\\
&&-\left.\left\lbrace\frac{3e^{-\hat{r}}}{\hat{r}}
\left(\frac{\hat{r}}{6}+\frac{1}{\hat{r}}+\frac{1}
{\hat{r}^2}+\frac{1}{2}\right)
-\frac{1}{\hat{r}^3}\right\rbrace\cos^2\theta_r\right].
\label{alpha_aniso}
\end{eqnarray}
Thus, the Coulomb potential in the presence of strong magnetic field is 
modified as 
\begin{eqnarray}
{\rm Re}~V_C(r,\theta _r;T,B)&=&-\frac{\alpha {m_D^q}^2}{\mu_{D}}
\left[\frac{e^{-\hat{r}}}{\hat{r}}\left(\frac{\hat{r}}{4}
+\frac{1}{\hat{r}}+\frac{1}{\hat{r}^2}+\frac{1}{2}
+\frac{\mu_D^2}{{m_D^q}^2}\right)
-\frac{1}{\hat{r}^3}-\frac{1}{12}+\frac{\mu_D^2}{{m_D^q}^2}
\right.\nonumber\\
&&-\left.\left\lbrace\frac{3e^{-\hat{r}}}{\hat{r}}
\left(\frac{\hat{r}}{6}+\frac{1}{\hat{r}}+\frac{1}
{\hat{r}^2}+\frac{1}{2}\right)
-\frac{1}{\hat{r}^3}\right\rbrace\cos^2\theta_r\right].
\end{eqnarray}
Similarly, the medium modification to the string part in \eqref{tot_pot} can be 
written as
\begin{eqnarray}
{\rm Re}~V_S(\mathbf{r};T,B)&=&-\frac{4\sigma}{(2\pi)^2}
\int\frac{d^3\textbf{k}~(e^{i\bf{k.r}}-1)}
{\mathbf{k}^2~(\mathbf{k}^2+{\mu_{D}}^2+\frac{{m_D^q}^2}{2}
	\cos{2\beta_n})},\nonumber\\
&=&-\frac{4\sigma}{(2\pi)^2}\int\frac
{d^3\textbf{k}~(e^{i\bf{k.r}}-1)}{\mathbf{k}^2(\mathbf{k}^2+
	{\mu_{D}}^2)}+\frac{2\sigma {m_D^q}^2}{(2\pi)^2}
\int \frac
{d^3\textbf{k}(e^{i\bf{k.r}}-1)~\cos2\beta_n}{\mathbf{k}^2(\mathbf{k}^2
	+{\mu_{D}}^2)^2},\nonumber\\
&\equiv &{\rm Re}~V_S^{(1)}(\mathbf{r},T,B)+{\rm Re}~V_S^{(2)}(\mathbf{r},T,B),
\end{eqnarray}
where, the first term in the string part is
\begin{eqnarray}
{\rm Re}~V_S^{(1)}(r;T,B)&=&-\frac{4\sigma}{(2\pi)^2}\int\frac
{d^3\textbf{k}~(e^{i\bf{k.r}}-1)}{\mathbf{k}^2(\mathbf{k}^2+{\mu_{D}}^2)},
\nonumber\\
&=&\frac{2}{\mu_D}\sigma\hat{r}
\left(\frac{e^{-\hat{r}}}{\hat{r}^2}-\frac{1}{\hat{r}^2}\right) +\frac{2\sigma}{\mu_{D}},
\label{string_iso}
\end{eqnarray} 
and using the same transformation (\ref{transfer}), the second term 
is calculated as
\begin{eqnarray}
{\rm Re}~V_S^{(2)}(r,\theta_r;T,B)&=&\frac{2\sigma 
	{m_D^q}^2}{(2 \pi)^2}\int{\frac{d^3\textbf{k}~(e^{i\bf{k.r}}-1)
		~(2\cos ^2{\beta _n}-1)}{\mathbf{k}^2(\mathbf{k}^2+{\mu_{D}}^2)^2}},\nonumber\\
&=&\frac{4 {m_D^q}^2 }{{\mu_{D}}^3}\sigma\hat{r}
\left[\frac{e^{-\hat{r}}}{\hat{r}}\left(\frac{1}{2\hat{r}}
+\frac{1}{\hat{r}^2}+\frac{1}{\hat{r}^3}+\frac{1}{8}\right)
+\frac{1}{24\hat{r}}-\frac{1}{\hat{r}^4}\right.\nonumber\\
&&-\left.\left\lbrace\frac{3e^{-\hat{r}}}{\hat{r}}
\left(\frac{5}{12\hat{r}}+\frac{1}{\hat{r}^2}+\frac{1}
{\hat{r}^3}+\frac{1}{12}\right)
+\frac{1}{12\hat{r}^2}-\frac{1}{\hat{r}^4}
\right\rbrace\cos^2\theta_r\right].
\label{string_aniso}
\end{eqnarray}
Thus the medium modification to the string part in the presence of strong
$\bf{B}$ becomes  
\begin{eqnarray}
{\rm Re}~V_S(r,\theta_r;T,B)&=&\frac{4 {m_D^q}^2 }{{\mu_{D}}^3}\sigma\hat{r}
\left[\frac{e^{-\hat{r}}}{\hat{r}}\left(\frac{1}{2\hat{r}}
+\frac{1}{\hat{r}^2}+\frac{1}{\hat{r}^3}+\frac{1}{8}
+\frac{\mu_D^2}{2\hat{r}{m_D^q}^2}\right)
+\frac{1}{24\hat{r}}-\frac{1}{\hat{r}^4}
-\frac{\mu_D^2}{2\hat{r}^2{m_D^q}^2}\right.\nonumber\\
&&+\left.\frac{\mu_D^2}
{2\hat{r}{m_D^q}^2}-\left\lbrace\frac{3e^{-\hat{r}}}{\hat{r}}
\left(\frac{5}{12\hat{r}}+\frac{1}{\hat{r}^2}+\frac{1}
{\hat{r}^3}+\frac{1}{12}\right)
+\frac{1}{12\hat{r}^2}-\frac{1}{\hat{r}^4}
\right\rbrace\cos^2\theta_r\right].
\end{eqnarray} 

So the real-part of the medium modified potential consists
of central and noncentral components
\begin{eqnarray*}
	{\rm Re}~V(r,\theta_r;T,B)&=& {\rm Re}~V_{{\rm central}}(r;T,B)+
	{\rm Re}~V_{{\rm noncentral}}(r,\theta_r;T,B),
\end{eqnarray*} 
where the central component is
\begin{eqnarray}
{\rm Re}~V_{{\rm central}}(r;T,B)&=&
-\frac{\alpha {m_D^q}^2}{\mu_{D}}
\left[\frac{e^{-\hat{r}}}{\hat{r}}\left(\frac{\hat{r}}{4}
+\frac{1}{\hat{r}}+\frac{1}{\hat{r}^2}+\frac{1}{2}
+\frac{\mu_D^2}{{m_D^q}^2}\right)
-\frac{1}{\hat{r}^3}-\frac{1}{12}+\frac{\mu_D^2}{{m_D^q}^2}
\right]\nonumber\\
&&+\frac{4 {m_D^q}^2 }{{\mu_{D}}^3}\sigma\hat{r}
\left[\frac{e^{-\hat{r}}}{\hat{r}}\left(\frac{1}{2\hat{r}}
+\frac{1}{\hat{r}^2}+\frac{1}{\hat{r}^3}+\frac{1}{8}
+\frac{\mu_D^2}{2\hat{r}{m_D^q}^2}\right)
+\frac{1}{24\hat{r}}-\frac{1}{\hat{r}^4}\right.\nonumber\\
&&-\left.\frac{\mu_D^2}{2\hat{r}^2{m_D^q}^2}+\frac{\mu_D^2}
{2\hat{r}{m_D^q}^2}\right]~,
\label{central}
\end{eqnarray}
and the noncentral component is
\begin{eqnarray}
{\rm Re}~V_{{\rm noncentral}}(r, \theta_r; T,B)&=&
\cos^2\theta_r\left[\frac{\alpha {m_D^q}^2}{\mu_{D}}
\left\lbrace\frac{3e^{-\hat{r}}}{\hat{r}}
\left(\frac{\hat{r}}{6}+\frac{1}{\hat{r}}+\frac{1}
{\hat{r}^2}+\frac{1}{2}\right)
-\frac{1}{\hat{r}^3}\right\rbrace\right.\nonumber\\
&&-\left.\frac{4 {m_D^q}^2 }{{\mu_{D}}^3}\sigma\hat{r}
\left\lbrace\frac{3e^{-\hat{r}}}{\hat{r}}
\left(\frac{5}{12\hat{r}}+\frac{1}{\hat{r}^2}+\frac{1}
{\hat{r}^3}+\frac{1}{12}\right)
+\frac{1}{12\hat{r}^2}-\frac{1}{\hat{r}^4}
\right\rbrace\right].
\label{noncentral}
\end{eqnarray}
%%%%%%%%%%%%%%%%%%%%%%%%%%%%%%%%%%%%%%%%%%%%%%%%%%%%%%%%%%%%%%%%%%
\begin{center}
\begin{figure}
\begin{tabular} {c c}
\includegraphics[width=8cm]{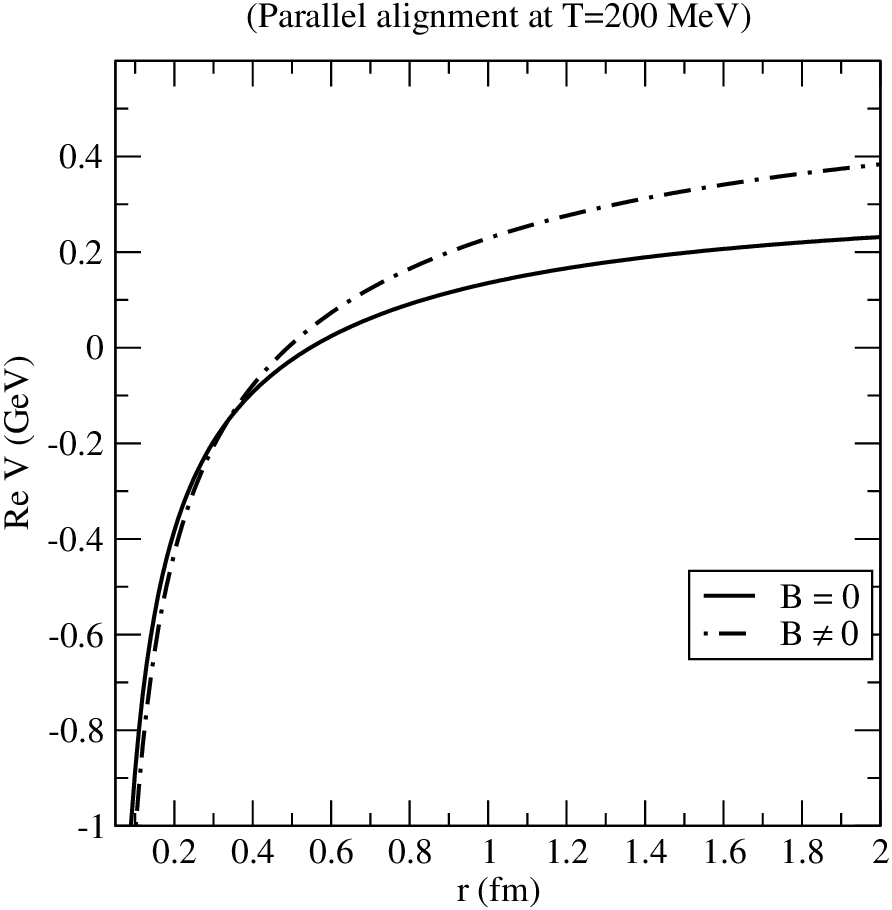}&
\hspace{1cm}
\includegraphics[width=8cm]{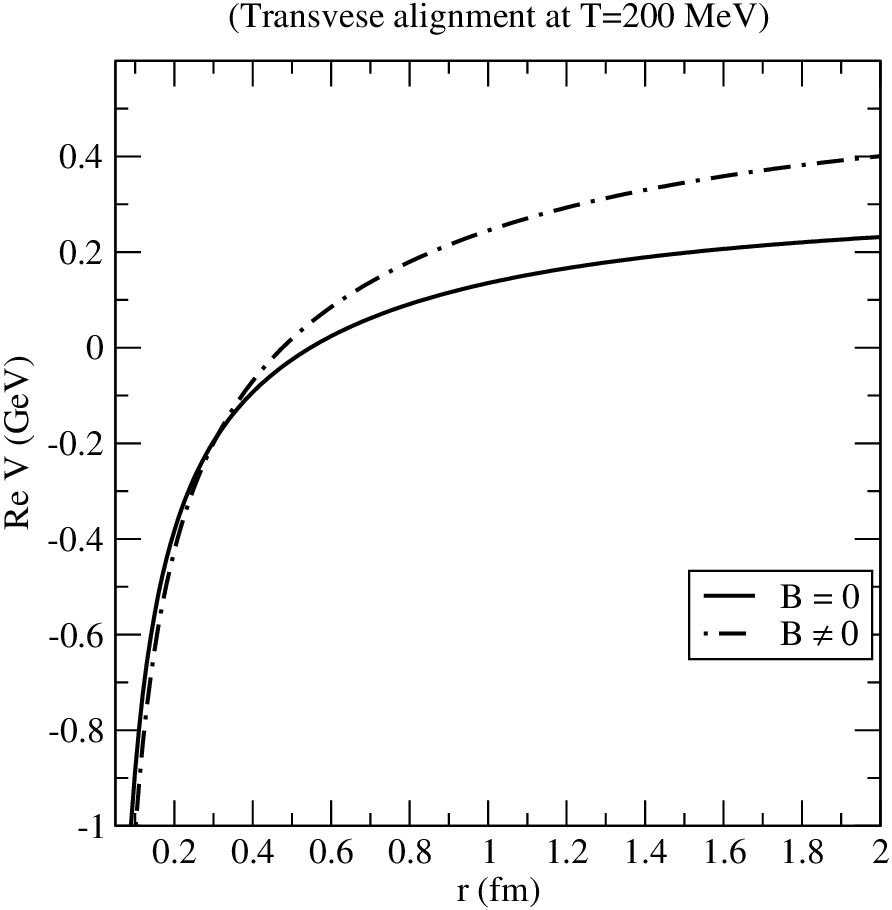}\\
(a) & (b)
\end{tabular}  
\caption{(a) Real-part of the potential in a hot QCD medium (at a temperature 
T=200 MeV) as a function of
inter-quark separation along the direction of strong magnetic field 
i) at ($eB=15 m_\pi^2$), ii) for $B=0$.
(b) Same as in (a) but the orientation becomes transverse.}
\label{parallel}
\end{figure}
\end{center}
%%%%%%%%%%%%%%%%%%%%%%%%%%%%%%%%%%%%%%%%%%%%%%%%%%%%%%%%%%%%%%%%%%
It is thus inferred that the strong magnetic field introduces angular
dependence into the $Q \bar Q$ interaction.
%, {\em i.e.} the interaction depends on the alignment of $Q \bar Q$ pair 
%with respect to the direction of strong magnetic field. 
To be specific, $Q\bar Q$ interaction is more attractive
when the $Q \bar Q$ pair is aligned transverse  to the magnetic field
than when the pair is aligned (parallel alignment) along the magnetic
field, which is reflected in Figure \ref{parallel}. 
%{(\color{red} yes it is right, it is due to the fact that, in 
%strong $\bf{B}$ the motion of the quark is restricted only
%in the longitudinal direction (in the direction of the B) so 
%there will be less no of quarks in the transverse region so effective 
%Debye mass will be less in transverse direction as compared to the 
%parallel direction (mostly gluons will contribute to the Debye mass 
%since there will be less no of quarks around a static charge))}.
We have observed that in the presence 
of the strong $\bf{B}$, the $Q\bar{Q}$
potential gets less screened compared its counterpart
in the absence of magnetic field,
which is due to softening of the screening/Debye masses.

\begin{center}
\begin{figure}
\begin{tabular} {c c}
\includegraphics[width=8cm]{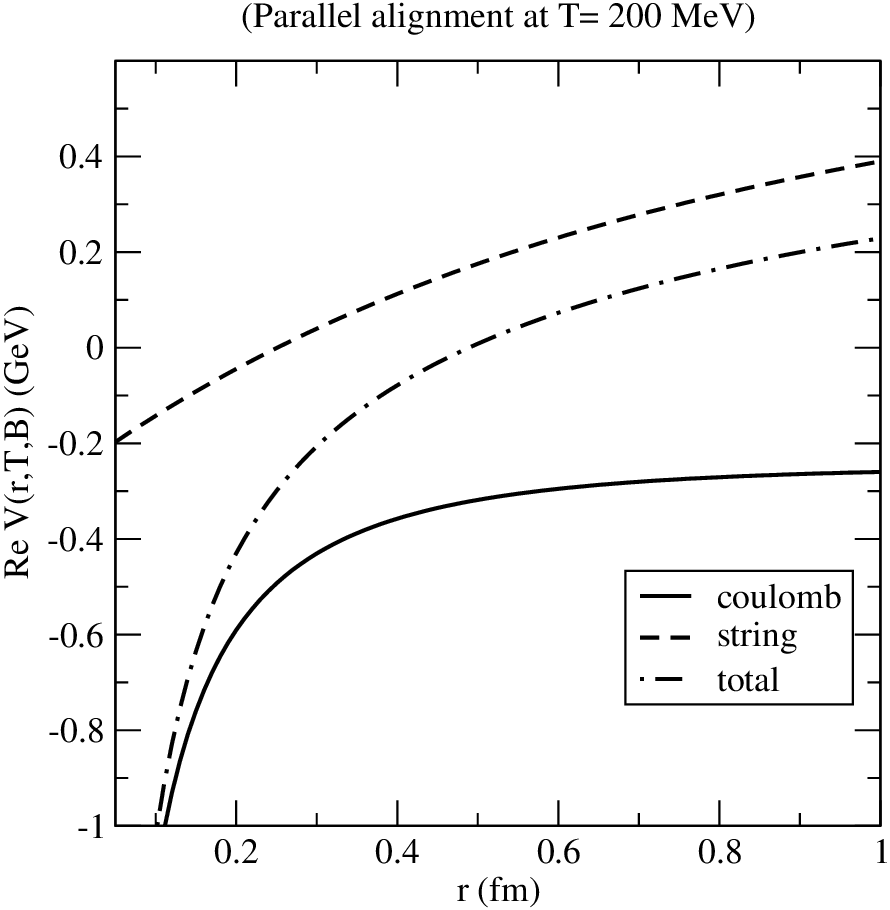}&
\hspace{1cm}
\includegraphics[width=8cm]{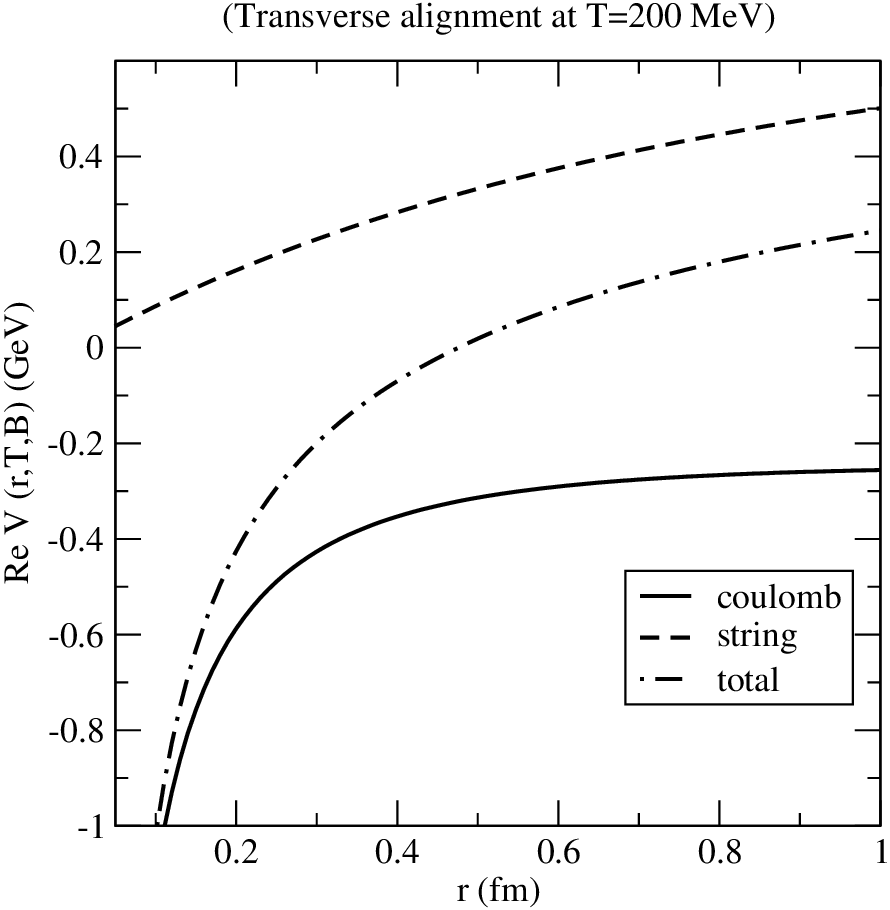}\\
(a) & (b)
\end{tabular}  
\caption{Decomposition of real-part of the potential in Figure 
1. (a) and (b) into the individual Coulomb, string components
as a function of inter-quark separation at T=200 MeV. For the
completeness, the sum total of individual components is also
displayed.}
\label{mag1}
\end{figure}
\end{center}

To decipher the effects of strong magnetic field on the (real) 
potential (in Figure 1) minutely, we postmortem it by decomposing 
into the Coulomb and string terms in Figure \ref{mag1}. We have
found that the (strong) magnetic field affects the string part more
than the coulomb part and the effect is more pronounced in the 
perpendicular alignment. The sting part increases (decreases) in the 
perpendicular (parallel) alignment whereas the coulomb part increases 
very slightly in both cases.
%%%%%%%%%%%%%%%%%%%%%%%%%%%%%%%%%%%%%%%%%%%%%%%%%%%%%%%%%%%%%%%%%%%%%%
%The angular dependence in heavy quark interaction was earlier
%revealed in the background of magnetic field by lattice studies~\cite{}.
The effects of a background magnetic field on 
the screening of both electric and magnetic fields in the deconfined 
medium were much earlier studied by computing the electric and magnetic 
electric screening masses, respectively, by measuring the Polyakov loop 
correlators on the lattice~\cite{Bonati:PRD89'2014,Bonati:PRD94'2016,
Bonati:PRD95'2017}. They found that the magnetic field enhances an increase of both 
screening masses and in addition, induces an anisotropy in Polyakov loop 
correlators, which in turn is translated into an anisotropy in
$Q \bar Q$ interacton. 
%Both screening masses are found to increase linearly 
%with the magnetic field and the influence of the magnetic field on the 
%two masses is enhanced at lower temperatures and is asymptotically 
%diminished in the higher temperature. Thus our aforesaid results on 
%the Debye mass qualitatively agree with their findings for the 
%electric screening mass, which is of interest to us for the screening 
%of the heavy quark potential. 
However, the lattice estimates for the electric screening masses are 
somehow much larger than our results, which may be due to
the large nonperturbative effects, beyond the scope of 
our perturbative framework.

%%%%%%%%%%%%%%%%%%%%%%%%%%%%%%%%%%%%%%%%%%%%%%
\subsubsection{Imaginary-part}
%%%%%%%%%%%%%%%%%%%%%%%%%%%%%%%%%%%%%%%%%%%%%%%
We had seen earlier in \eqref{ima_per} that the imaginary-part of permittivity is 
separable into the quark- and gluon-loop contribution. So we first find out 
the quark-loop (labelled as $q$) contribution \eqref{quark_per} to the 
imaginary-part of the potential, which, however, vanishes
\begin{eqnarray}
\rm{Im}~V^q(\mathbf{r};T,B)=0.
\end{eqnarray}
This happens due to the appearance of Dirac delta function, $\delta (\rm k_z)$ in 
the  $q$-contribution to the imaginary part of permittivity. 
The appearance of delta function can be understood from the constraint on the 
motion of the quarks in LLL states, due to the strong magnetic field ($z$
direction). As a subsequent consequence, the dispersion relation for the 
massless quarks in the LLL states will simply be, $\omega=\pm \rm k_z$. 
So, in the static limit, there will be no longitudinal energy-momentum 
transfer in the inelastic process involving massless quarks.

Next we will calculate the gluon-loop contribution of the 
permittivity \eqref{gluon_per}, where the Coulomb term is given by 
\begin{eqnarray}
{\rm Im}~V_C^g(\mathbf{r};T,B)&=&\frac{1}{(2\pi)^{3/2}}\int {d^3\textbf{k}}~
(e^{i\bf{k.r}}-1)\left(-\sqrt{\frac{2}{\pi}}\frac{\alpha}{\mathbf{k}^2}\right)
\left(\frac{-{m_D^g}^2 \pi T \mathbf{k}^2}
{{\rm k}(\mathbf{k}^2
+{m_D^g}^2+{m_D^q}^2\cos^2\beta_n)^2 }\right),\nonumber\\
&=&\frac{\alpha {m_D^g}^2 T}{2\pi}\int\frac{d^3\textbf{k}~(e^{i\bf{k.r}}-1)}
{{\rm k}(\mathbf{k}^2+
{\mu_{D}}^2)^2}-\frac{\alpha {m_D^g}^2{m_D^q}^2 T}{2\pi}\int \frac
{d^3\textbf{k}(e^{i\bf{k.r}}-1)~\cos2\beta_n}{{\rm k}(\mathbf{k}^2+{\mu_{D}}^2)^3},
\nonumber\\
&\equiv &\Psi_ 1(\hat{r})+\Psi_ 2(\hat{r},\theta_r),
\end{eqnarray}
where $\Psi_ 1$ is given by
\begin{eqnarray}
\Psi_ 1(\hat{r})=\frac{-\alpha {m_D^g}^2 T}{{\mu_{D}}^2} \phi_ 0(\hat{r}).
\label{alpha_iso1}
\end{eqnarray}
%with 
%\begin{eqnarray}
%\phi_ 0(\hat{r})=\frac{-\hat{r}^2}{9}(-4+3\gamma _ E+3\log{\hat{r}}),
%\end{eqnarray}
%and $\gamma_ E$ is the Euler constant.
 By substituting the transformation 
between the angular variables
in momentum-space anisotropy and the coordinate-space anisotropy \eqref{transfer}, 
$\Psi_ 2$ is obtained as 
\begin{eqnarray}
\Psi_ 2(\hat{r},\theta_ r) &=&-\frac{2\alpha {m_D^g}^2{m_D^q}^2 T}{{\mu_{D}}^4}
\left[\int_ 0^{\infty} 
\frac{zdz}{(z^2+1)^3}\left\{\left(-\frac{\sin{(z\hat{r})}}{(z\hat{r})}-
\frac{2\cos{(z\hat{r})}}{(z\hat{r})^2}+\frac{2\sin{(z\hat{r})}}
{(z\hat{r})^3}\right)\right.\right. \nonumber\\
&&+ \left.\left.\left(\frac{2\sin{(z\hat{r})}}{(z\hat{r})}+
\frac{6\cos{(z\hat{r})}}{(z\hat{r})^2}-\frac{6\sin{(z\hat{r})}}
{(z\hat{r})^3}\right )\cos^ 2{\theta_ r}\right\}
 +\frac{1}{3}\int_ 0^{\infty} 
\frac{zdz}{(z^2+1)^3}\right].
\label{alpha_aniso1}
\end{eqnarray}
which, in addition to $r$, also depends on the relative orientation of 
$Q \bar Q$ pair  with respect to the magnetic field.

\par
Similarly the imaginary-part to the string term is obtained as
%the gluon-loop permittivity \eqref{gluon_per} 
\begin{eqnarray}
{\rm Im}~V_S^g(\mathbf{r};T,B)&=&\frac{1}{(2\pi)^{3/2}}\int {d^3\textbf{k}}~
(e^{i\bf{k.r}}-1)\left(-\frac{4\sigma}{\sqrt{2\pi}\mathbf{k}^4}\right)
\left(\frac{-{m_D^g}^2 \pi T \mathbf{k}^2}
{{\rm k}(\mathbf{k}^2
+{m_D^g}^2+{m_D^q}^2\cos^2\beta_n)^2 }\right),\nonumber\\
&\equiv &\Psi_3 (\hat{r})+\Psi_4 (\hat{r},\theta_ r).
\end{eqnarray}
where $\Psi_3 (\hat{r})$ is calculated as 
\begin{eqnarray}
\Psi_ 3(\hat{r})&=&\frac{-2\sigma {m_D^g}^2 T}{{\mu_{D}}^4}\Psi_ 0(\hat{r}),
\label{string_iso1}
\end{eqnarray}
%with
%\begin{eqnarray}
%\phi_ 1(\hat{r})=\frac{\hat{r}^2}{6}+\frac{(-107+60
%\gamma_ E+60\log{\hat{r}})\hat{r}^4}{3600}.
%\end{eqnarray}
and $\Psi_4$ is obtained as
\begin{eqnarray}
\Psi_4 (\hat{r},\theta_ r)&=&
-\frac{4\sigma {m_D^g}^2{m_D^q}^2 T}{{\mu_{D}}^6}\left[\int_ 0^{\infty} 
\frac{dz}{z(z^2+1)^3}\left\{\left(-\frac{\sin{(z\hat{r})}}{(z\hat{r})}-
\frac{2\cos{(z\hat{r})}}{(z\hat{r})^2}+\frac{2\sin{(z\hat{r})}}
{(z\hat{r})^3}\right)\right.\right.\nonumber\\
&&+ \left. \left.\left(\frac{2\sin{(z\hat{r})}}{(z\hat{r})}+
\frac{6\cos{(z\hat{r})}}{(z\hat{r})^2}-\frac{6\sin{(z\hat{r})}}
{(z\hat{r})^3}\right )\cos^ 2{\theta_ r}\right\}+
 \frac{1}{3}\int_ 0^{\infty} \frac{dz}{z(z^2+1)^3}\right].
 \label{string_aniso1}
\end{eqnarray}

Thus, {\em like} the real-part, the imaginary-part of the potential also 
consists of central and noncentral components
\begin{eqnarray*}
 {\rm Im}~V(r,\theta_r;T,B)&=& {\rm Im}~V_{{\rm central}}(r;T,B)+
  {\rm Im}~V_{{\rm noncentral}}(r,\theta_r;T,B),
\end{eqnarray*}
where the central component is written as
{\small
\begin{eqnarray}
{\rm Im}~V_{{\rm central}}(r;T,B)&=&
\frac{\alpha {m_D^g}^2 T\hat{r}^2}{9{\mu_{D}}^2}
(-4+3\gamma_ E+3\log{\hat{r}})\nonumber\\
&&-\frac{2\sigma {m_D^g}^2 T}{{\mu_{D}}^4}
\left(\frac{\hat{r}^2}{6}+\frac{(-107+60
\gamma_ E+60\log{\hat{r}})\hat{r}^4}{3600}\right)\nonumber\\
&&-\frac{2\alpha {m_D^g}^2{m_D^q}^2 T}
{{\mu_{D}}^4}\left[\int_ 0^{\infty} 
\frac{zdz}{(z^2+1)^3}\left(-\frac{\sin{(z\hat{r})}}{(z\hat{r})}-
\frac{2\cos{(z\hat{r})}}{(z\hat{r})^2}
 +\frac{2\sin{(z\hat{r})}}
{(z\hat{r})^3}+\frac{1}{3}\right)\right]\nonumber\\
&&-\frac{4\sigma {m_D^g}^2{m_D^q}^2 T}{{\mu_{D}}^6}\left[\int_ 0^{\infty} 
\frac{dz}{z(z^2+1)^3}\left(-\frac{\sin{(z\hat{r})}}{(z\hat{r})}
-\frac{2\cos{(z\hat{r})}}{(z\hat{r})^2}+\frac{2\sin{(z\hat{r})}}
{(z\hat{r})^3}+\frac{1}{3}\right)\right],
\end{eqnarray}}
while the noncentral component is written as 
{\small
\begin{eqnarray}
{\rm Im}~V_{{\rm noncentral}}(r,\theta_r;T,B)&=&
\left[-\frac{2\alpha {m_D^g}^2{m_D^q}^2 T}
{{\mu_{D}}^4} \left\{ \int_ 0^{\infty} 
\frac{zdz}{(z^2+1)^3} \left(\frac{2\sin{(z\hat{r})}}{(z\hat{r})}+
\frac{6\cos{(z\hat{r})}}{(z\hat{r})^2} -\frac{6\sin{(z\hat{r})}}
{(z\hat{r})^3}\right )\right\}\right. \nonumber \\
&&\left. -\frac{4\sigma {m_D^g}^2{m_D^q}^2 T}{{\mu_{D}}^6}\left\{\int_ 0^{\infty} 
\frac{dz}{z(z^2+1)^3}\left(\frac{2\sin{(z\hat{r})}}
{(z\hat{r})} +\frac{6\cos{(z\hat{r})}}{(z\hat{r})^2}-\frac{6\sin{(z\hat{r})}}
{(z\hat{r})^3}\right )\right\} \right]\cos^ 2{\theta_ r}.
\end{eqnarray}}
%Thus, the strong magnetic field also conceives an angular dependence into 
%the imaginary-part of $Q \bar Q$ interaction.
\begin{center}
\begin{figure}
\begin{tabular} {c c}
\includegraphics[width=8cm]{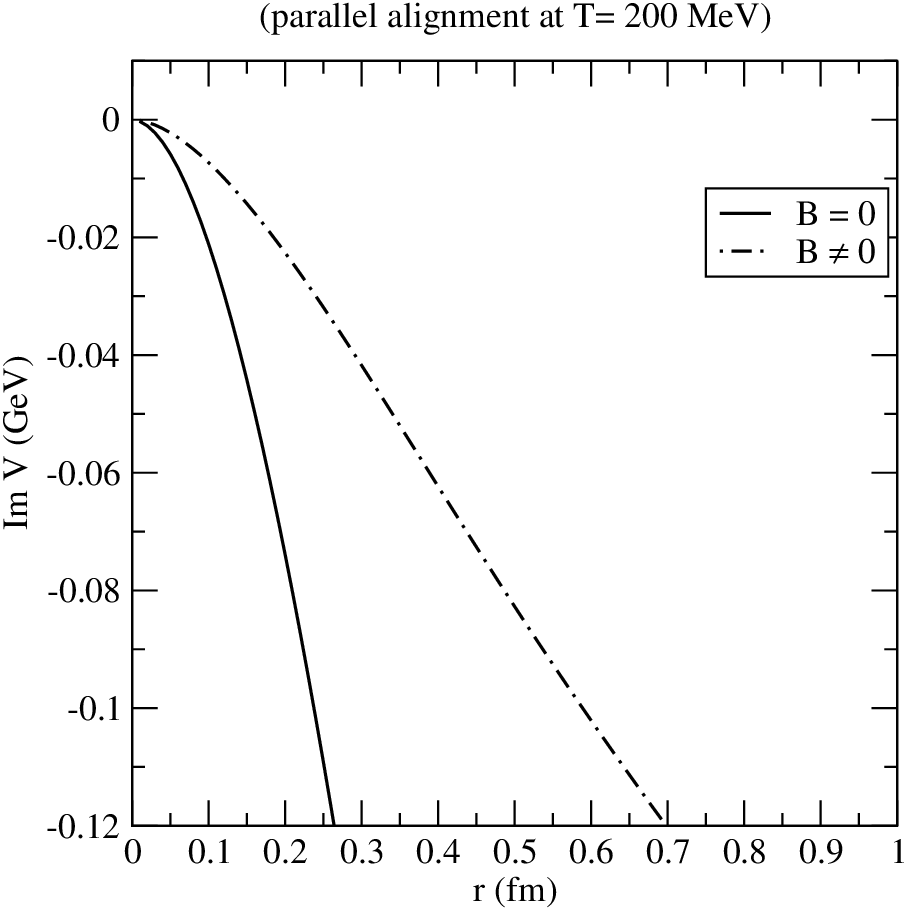}&
\hspace{1cm}
\includegraphics[width=8cm]{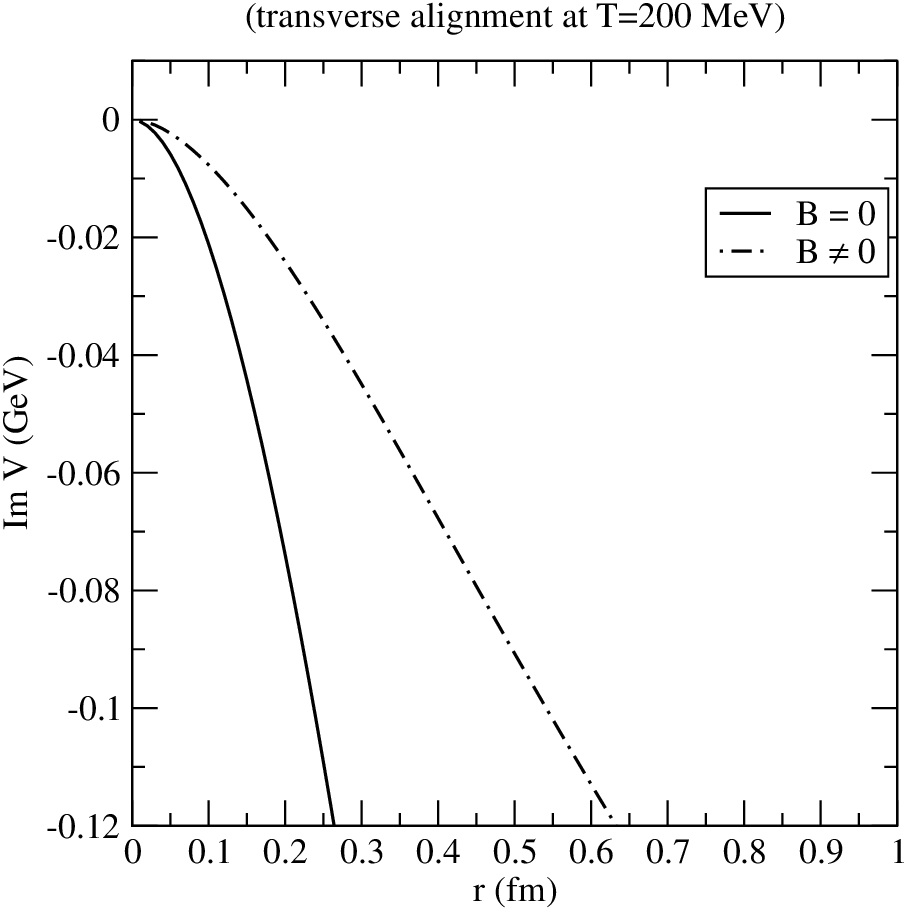}\\
(a) & (b)
\end{tabular}  
\caption{Variation of the imaginary part of the potential as a function
of $r$ at T=200 MeV with the identical of Figure 1}
\label{ima_para}
\end{figure}
\end{center}
We have now displayed the effect of strong magnetic field on the 
imaginary part of the heavy quark potential in Figure \ref{ima_para} 
as a function of interparticle separation ($r$) with respect to
the direction of magnetic field for two orientations
of ${\bf r}$ with respect to the direction of magnetic field (direction
of anisotropy itself). It is found that the magnitude of
imaginary-part in general gets reduced in strong $B$ compared to 
$B=0$, which can again be
attributed due to the softening of the Debye mass. However, the
decrease (in magnitude) is lesser in the transverse direction than
in the direction of magnetic field.
%%%%%%%%%%%%%%%%%%%%%%%%%%%%%%%%%%%%%%%%%%%%%%%%%%%%%%%%%%%%%%%%%%%%
\section{Properties of Quarkonia in strong $B$ and its dissociation}
%%%%%%%%%%%%%%%%%%%%%%%%%%%%%%%%%%%%%%%%%%%%%%%%%%%%%%%%%%%%%%%%%%%%
In order to study how the presence of an external strong magnetic field 
affects the in-medium properties of $Q \bar Q$ (nonrelativistic)
bound states immersed in a hot QCD medium, we have solved the 
Schrodinger equation numerically with the potential thus obtained
in \eqref{central} and \eqref{noncentral}. Since the potential
is complex so the real- and imaginary-parts of the potential yield the binding
energies and in-medium widths of the bound states in the presence of 
strong $\bf{B}$, respectively. Since real-part has both spherical and 
nonspherical component and the nonspherical (angular) component is very 
small compared to the spherical component, so we have treated the 
nonspherical component as a perturbation and calculated 
the binding energies for the $J/\Psi$ and $\Upsilon$ states in
a first-order perturbation theory. The binding energies thus obtained
numerically decreases with the temperature (seen in Figure 4), however,
its value gets enhanced in comparison to the absence of $\bf{B}$ value. 

We have also calculated the in-medium widths $(\Gamma)$ of quarkonia 
with the imaginary part of the $Q\bar{Q}$
potential in first-order perturbation theory. Assuming the ground states
of $c \bar c$ and $b \bar b$ states (which are $J/\psi$ and $\Upsilon$,
respectively) as the Coulombic bound states, the widths are
calculated numerically from the relation: 
\begin{eqnarray}
\Gamma=-\int d^3 \mathbf{r}~|\Psi (r)|^2~ \text{Im}V(r,\theta_ r;T,B), 
\end{eqnarray}
where $\Psi(r)$ is the ground state wave function and is given by
\begin{eqnarray}
\Psi(r)=\frac{1}{\sqrt{\pi a_0^3}}e^{-\frac{r}{a_0}}
\end{eqnarray}
wth $a_0$ as the radius of the first Bohr orbit for $Q \bar Q$
bound state. The widths, $\Gamma$'s are found to increase with the 
temperature (Figure 4), but its magnitude gets reduced 
compared to $\bf{B}=0$ result.

Finally we have studied the quasi-free dissociation of quarkonium 
states by the competition
of binding energies and medium widths, which originate from the 
real- and imaginary-parts of the potential in a medium, respectively. 
The quantitative study of the dissociation is made by the above 
competition between screening and Landau damping, in particular, when the 
binding energy of a particular 
quarkonium state ($i$) is equal to the half of its width,
{\em i.e.} ${\rm B.E.|}_i =\frac{\Gamma_i}{2}$.
Since both quantities (B.E. and $\Gamma$) depend on the temperature and 
the strength of the strong magnetic field, so the above relation gives
the temperature of the hot medium ($T_d$) in a strong $B$ at which the 
$Q \bar Q$ state gets excited and moves to the continuum. We have
obtained the $T_d$'s for $J/\Psi$ and $\Upsilon$ as $1.59~\rm{T_C}$ and 
$2.22~\rm{T_C}$, respectively.
%{\large \color{red} try to give comparative results for B E, 
%width, dissociation temperature in isotropic, expansion driven anisotropy
%with specific anisotropic paramater (at least two values), $\xi$, magnetic field driven
%anisotropy with same value of $xi$}.
\begin{center}
\begin{figure}
\begin{tabular} {c c}
\includegraphics[width=8cm]{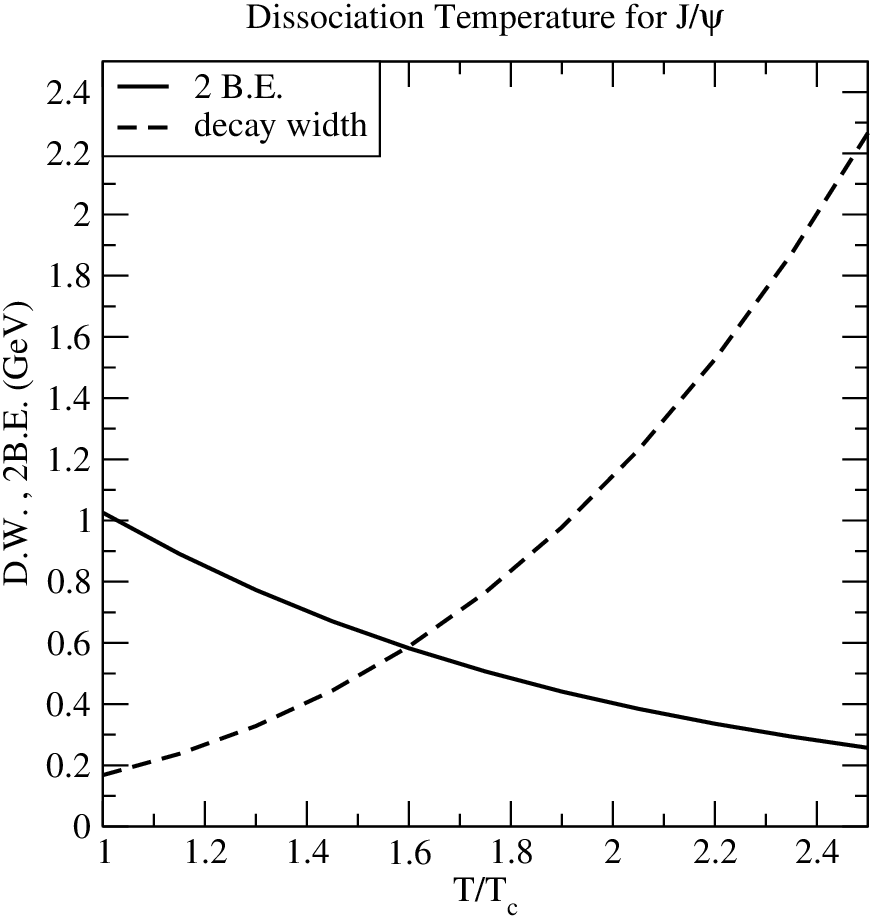}&
\hspace{1cm}
\includegraphics[width=8cm]{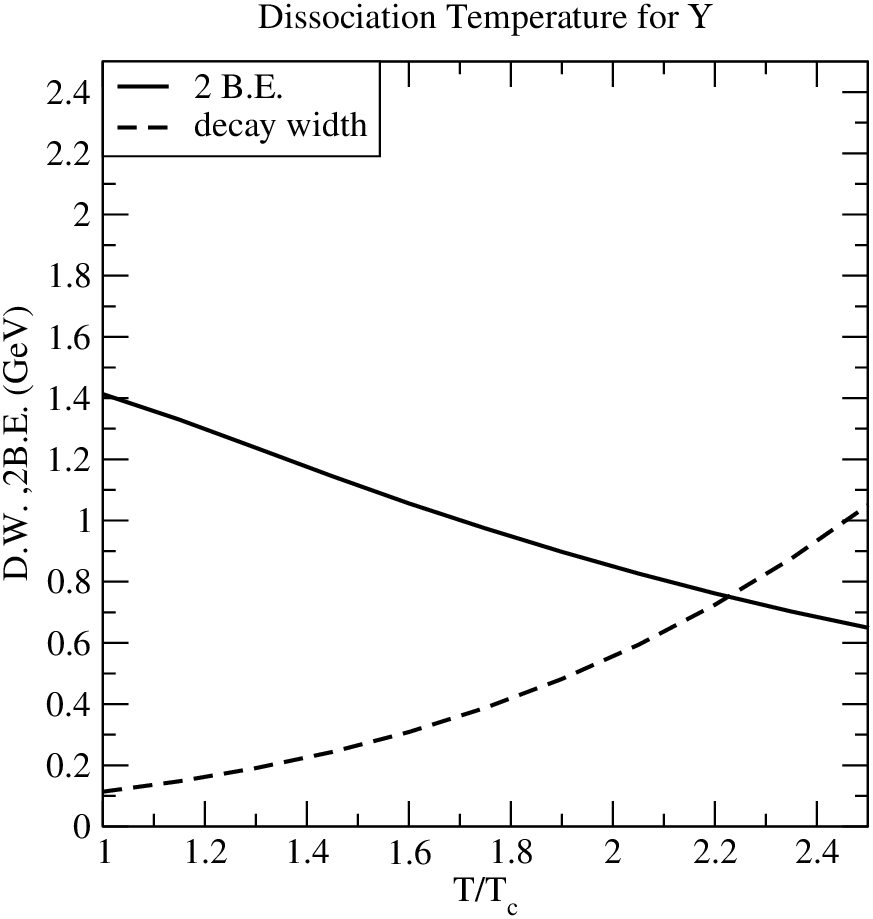}\\
(a) & (b)
\end{tabular}  
\caption{The decay width vs $2\times$ binding energy  for
 $J/\Psi$ and 
$\Upsilon$ in strong magnetic 
field  at $eB=15 m_{\pi}^2$.
}
\label{dissociation}
\end{figure}
\end{center}

%\begin{table}[h]
%\begin{center} 
%\caption{Dissociation temperature for $J/\psi$ and 
%$\Upsilon$ ground state in multiple of $T_c$}
%\begin{tabular}{|c|c|c|c|}
%\hline 
%Quarkonium &  Isotropic &
%   Magnetic field driven \\ 
%states & thermal medium  & anisotropic medium\\
%   &          & ($eB= 15 m_{\pi}^2$)\\
%\hline
% $J/\psi$ & 1.38  & 1.59 \\ 
%\hline
%$\Upsilon$ & 1.70  & 2.22 \\
%\hline 
%\end{tabular}
%\end{center} 
%\end{table}

\section{Conclusions and future outlook}
In this work, we have have delved into the effects of strong magnetic field on 
the ground state properties of $J/\psi$ and $\Upsilon$ in a hot QCD medium
through the color screening and the Landau damping phenomena.
% which are digitalized by the real- and imaginary-parts 
%of $Q \bar Q$ potential, respectively. 
For that purpose, we first thermalize the Schwinger propagator in the 
lowest Landau level (LLL) and the Feynman propagator for quarks and gluons, 
respectively to obtain the gluon self-energy for a thermal QCD medium 
with massless flavours in a strong magnetic field. We found that 
the medium does not contribute to the quark-loop contribution
rather its vacuum contribution yields an angular dependence to 
the self-energy. This finding can be envisaged by the equivalence 
between the massless QED in (1+1) dimension (Schwinger model) with the massless 
thermal QCD in strong magnetic field, which forbids any medium (finite 
temperature) correction to the self-energy.
Thus the self-energy introduces the angular dependence in the resummed 
propagators and, hence the permittivity of the medium becomes anisotropic, 
{\em i.e.} behaves like a tensor, which in fact inserts the 
nonspherical (anisotropic) term in $Q\bar{Q}$ potential. 

Overall the real part of the potential in the strong $\bf{B}$ 
is found more attractive as compared to the thermal
medium in the absence of magnetic field ($\bf{B}=0$), 
due to the softening of the Debye masses. Moreover, 
the $Q\bar{Q}$ potential is more attractive in the transverse alignment as 
compared to the parallel alignment of the $Q\bar{Q}$ pair with 
respect to the magnetic field, which is also seen in the lattice studies
~\cite{Bonati:PRD89'2014,Bonati:PRD94'2016}. On the other hand, the magnitude of the 
imaginary-part decreases, compared to $\bf{B}=0$ case. However, this
decrease is less pronounced in the transverse direction than the parallel 
alignment.

Finally we have solved the Schrodinger equation with the spherical
part of the (real) potential numerically to obtain the eigen function,
which in turn is used to calculate the first-order correction due to the 
nonspherical part of (real) of the potential in the time-independent 
perturbation theory. 
The binding energies for  $J/\Psi$ and $\Upsilon$ thus obtained are
found larger in comparison to the $\bf{B}=0$ case. Similarly we have also 
studied numerically the effect of the magnetic field 
on the medium induced width of quarkonia from the imaginary part of the 
potential in first-order perturbation theory, which, on the
contrary is smaller than in the absence of $\bf{B}$. Finally, with these
inputs on the properties of quarkonia in strong $B$, we have studied 
the quasi-free dissociation of $J/\psi$ and $\Upsilon$
in the magnetized thermal QCD medium with an optimized criterion
on the binding energy and width of a particular resonance - B.E.= Width 
($\Gamma$)/2. The dissociation temperatures ($T_d$) are thus 
found as $1.59T_c$ and $2.22T_c$, respectively, which are
larger than the $T_d$'s in the absence of $B$. Thus the presence
of strong magnetic field does not favour the early dissolution
of quarkonia in the medium.

The nonspherical (anisotropic) interaction in the potential 
could have consequences on the meson spectrum 
in heavy-ion phenomenology in the
meson spectrum~\cite{
Frasca:JHEP11'2013,Dudal:JHEP03'2013,Kerbikov:PRD87'2013,
Machado:PRD88'2013,Machado:PRD89'2014} because the perturbation
to the energy levels due to the nonspherical interaction 
may modify their production as well as decay rates etc. 
As we have noticed that the strong magnetic field affects the 
string part more than the Coulomb part, {\em especially} for
the perpendicular alignment of $Q \bar Q$ pairs. One
of the possible consequence is that the particle production,
mainly mesons through the strong breaking could be affected.
One of the corollary  of the anisotropic interaction 
may affect the thermalization process, which could be verified
through the measurement of the elliptic flow.

\section*{Acknowledgements}
We are thankful to Shubhalaxmi Rath for taking part in the
discussion on this work.
BKP is thankful to the CSIR (Grant No.03 (1407)/17/EMR-II),
Government of India for the financial assistance. 

\appendix
\appendixpage
\addappheadtotoc
\begin{appendices}
\renewcommand{\theequation}{A.\arabic{equation}}
\section{Anisotropic contribution in gluon self-energy}\label{anisotropy}

In a magnetic field, the energy levels of a quark ($f$) in vacuum get 
discretized into Landau ($n= 0,1,2, \cdots $) levels as
\begin{eqnarray}
\omega _{f,n}(p_L)=\sqrt{p_L^2+m_f^2+2n|q_fB|}.
\end{eqnarray} 
However, if the magnetic field is strong enough ($|q_fB| >>T^2$), quarks are 
confined to be in the LLL ($n=0$, due to the large energy gap ($\sim 
O(\sqrt{eB})$) between LLL and higher Landau levels ($n=1,2, \cdots$)
and results an anisotropy ($p_L << p_T$) in the momentum distribution of quarks 
with a negative anisotropic parameter $\xi= \frac{p_T^2}{2p_L^2}-1$).
Therefore, the distribution functions for the quarks $ n_F (p_0)$ 
in \eqref{quark_prop} can be approximated by the isotropic one, at least, for 
weak anisotropy ($\xi \ll 1$)
\begin{eqnarray}
n_F^{\rm aniso} (p)=\frac{1}{e^{\beta \sqrt{p^2-\xi(\bf{p.n})^2+m_f^2}}+1},
\label{disfun}
\end{eqnarray}
where, $p=(0,0,p_z)$ and $\bf{n}$ is the direction of the anisotropy, 
{\em i.e.} the direction of magnetic field.  For weak
anisotropy, we may expand the distribution function 
(\ref{disfun}) in the powers of $\xi$ and retain the term linear in 
$\xi$ only, 
\begin{eqnarray}
n_F^{\rm aniso} (p_0)=n_F(p_0)+\xi 
\frac{(\mathbf{p.n})^2}{2p_0 T} e^{\frac{|p_0|}{T}} n_F^2(p_0) + \cdots
\label{anfun1}
\end{eqnarray}
Therefore, the quark-loop contribution \eqref{quark_self} in a strong magnetic field
deviates from the same in isotropic medium \eqref{self_B} and 
is decomposed into isotropic and anisotropic components, 
\begin{eqnarray}
\Pi_{11}^{q,\mu\nu}(\rm k_{\parallel})&=&\Pi_{11(\rm iso)} ^{q,\mu\nu} 
(\rm k_{\parallel})+ \Pi_{11(\rm aniso)} ^{q,\mu\nu}(k_{\parallel}).
\label{self_tot}
\end{eqnarray}
We are now going to calculate the anisotropic
contribution up to the order $\xi$ 
\begin{eqnarray}
\Pi_{11(\rm aniso)}^{q,\mu\nu}(k_{\parallel})= \xi \left( 
I_1^{\mu\nu}(k_{\parallel})+I_2^{\mu\nu}(k_{\parallel})
+I_3^{\mu\nu}(k_{\parallel})+I_4^{\mu\nu}(k_{\parallel})
\right),
\label{ima_aniso}
\end{eqnarray}
where $I^{\mu \nu}_1$, $I^{\mu \nu}_2$, $I^{\mu \nu}_3$ and 
$I^{\mu \nu}_4$ are given by
\begin{eqnarray}\label{aniso1}
I_1^{\mu\nu}(\rm k_{\parallel})&=&-\frac{g'^2}{2}\int 
\frac{dp_{\parallel}^2}{(2\pi)^4}~
L^{\mu\nu}\left(\frac{2 \pi ~ n_F^2(p_0)~\exp{(\frac{|p_0|}{T})}
~(\mathbf{p.n})^2}{2p_0 T(q_{\parallel}^2-m_f^2+i\epsilon)}\right)
\delta(p_{\parallel}^2-m_f^2),\\
\label{aniso2}
I_2^{\mu\nu}(\rm k_{\parallel})&=&-\frac{g'^2}{2}\int 
\frac{dp_{\parallel}^2}{(2\pi)^4}~
L^{\mu\nu}\left(\frac{2 \pi~ n_F^2(p_0)~\exp{(\frac{|p_0|}{T})}
~(\mathbf{p.n})^2 ~2\pi i n_F(q_0)}{2p_0 T}\right)
\delta(p_{\parallel}^2-m_f^2) \nonumber \\
&& \times \delta(q_{\parallel}^2-m_f^2),\\
\label{aniso3}
I_3^{\mu\nu}(\rm k_{\parallel})&=&-\frac{g'^2}{2}\int \frac{dp_{\parallel}^2}{(2\pi)^4}~
L^{\mu\nu}\left(\frac{2 \pi ~ n_F^2(q_0)~\exp{(\frac{|q_0|}{T})}
~(\mathbf{q.n})^2}{2q_0 T(p_{\parallel}^2-m_f^2+i\epsilon)}\right)
\delta(q_{\parallel}^2-m_f^2),\\
\label{aniso4}
I_4^{\mu\nu}(\rm k_{\parallel})&=&-\frac{g'^2}{2}\int \frac{dp_{\parallel}^2}{(2\pi)^4}~
L^{\mu\nu}\left(\frac{2 \pi ~ n_F^2(q_0)~\exp{(\frac{|q_0|}{T})}
~(\mathbf{q.n})^2~ 2\pi i n_F(p_0)}{2q_0 T}\right)
\delta(q_{\parallel}^2-m_f^2) \nonumber \\
&& \times \delta(p_{\parallel}^2-m_f^2). 
\end{eqnarray}
The longitudinal component of the real-part of the anisotropic component 
\eqref{self_tot} comes out to be zero
\begin{eqnarray}
\text{Re}\Pi_{11(\rm aniso)}^{q,\parallel}(\rm k_z)=0,
\label{self_aniso}
\end{eqnarray}
because
\begin{eqnarray*}
\text{Re}~I_1^{q,\parallel}({\rm k_z}) + \text{Re}~I_3^{q,\parallel}({\rm k_z})
&=&0, \\
\text{Re}~I_2^{q,\parallel}({\rm k_z})=0,\quad
\text{Re}~I_4^{q,\parallel}({\rm k_z})=0.
\end{eqnarray*}

Similarly the imaginary-part of the anisotropic contribution also vanishes
\begin{eqnarray}
\text{Im}\Pi^{q,\parallel}_{11(\rm aniso)} {(\rm k_z)}=0.
\end{eqnarray}
\end{appendices}

\end{document}